\documentclass[sigplan,nonacm]{acmart} 

\usepackage{subcaption}
\usepackage{kotex}
\usepackage{booktabs,pifont}
\usepackage{xspace}
\usepackage{multirow}
\usepackage{cuted}

\graphicspath{{figures/}}

\begin{document}


\title[Efficient Expert-Parallel Communication on PCIe-Connected Consumer GPUs]
{Efficient Expert-Parallel Communication \\ on PCIe-Connected Consumer GPUs}

\author{Jaehwan Lee}
\affiliation{%
  \department{Dept. of Computer Science and Engineering}
  \institution{Seoul National University}
  \city{}
  \country{}
}
\email{jaehwan@aces.snu.ac.kr}
\author{Sangmin Lee}
\affiliation{%
  \department{Dept. of Computer Science and Engineering}
  \institution{Seoul National University}
   \city{}
  \country{}
}
\email{sangmin.lee@snu.ac.kr}
\author{Chaewon Kim}
\affiliation{%
  \department{Dept. of Computer Science and Engineering}
  \institution{Seoul National University}
   \city{}
  \country{}
}
\email{chaewon@aces.snu.ac.kr}
\author{Junsik Shin}
\affiliation{%
  \department{Dept. of Computer Science and Engineering}
  \institution{Seoul National University}
   \city{}
  \country{}
}
\email{junsik@aces.snu.ac.kr}
\author{Jaejin Lee}
\affiliation{%
  \department{Dept. of Computer Science and Engineering}
  \department{Graduate School of Data Science}
  \institution{Seoul National University}
   \city{}
  \country{}
}
\email{jaejin@snu.ac.kr}
\renewcommand{\shortauthors}{Lee et al.}

\begin{abstract}
Expert parallelism (EP) enables inference of large
Mixture-of-Experts (MoE) models by placing their experts across
multiple GPUs, but requires substantial communication between GPUs at every MoE layer. As contemporary MoE models activate more experts per token,
this communication accounts for a growing fraction of inference time.
The cost becomes particularly pronounced on PCIe-based consumer GPU systems, where all inter-GPU transfers traverse CPU memory. 
However, existing MoE-specialized EP communication libraries assume that direct GPU-to-GPU access is available, largely overlooking consumer GPUs. 
Therefore, most LLM frameworks instead rely on NCCL, whose CPU-staged
communication incurs redundant PCIe transfers and competes with expert
computation for GPU resources, limiting their overlap.
We present ThunderEP, a novel communication design for such systems that removes the relay hops of traditional ring algorithm, moves data through DMA engines to avoid compute resource contention, and minimizes synchronization latency by reducing the polling overhead of completion flags in CPU memory. 
We integrate the proposed design into vLLM and evaluate it on three widely used MoE models.
Experiments on two PCIe systems equipped with RTX 4090 and RTX 5090 GPUs show that ThunderEP achieves average speedups of 2.00$\times$ and 1.53$\times$ over NCCL for dispatch and combine, respectively, and up to 1.66$\times$ end-to-end speedup over state-of-the-art MoE inference frameworks.
\end{abstract}

\maketitle 

\section{Introduction}
\label{sec:intro}







Mixture-of-Experts (MoE)~\cite{jacobs1991adaptive} has become a standard architecture for scaling
frontier large language models (LLMs)~\cite{team2026kimi,xu2026deepseek-v4.1-flash,zeng2026glm}. The expert parameters of
modern MoE-based LLMs, however, are often too large to reside on a single GPU~\cite{liu2024deepseek-v3}.
Serving these models therefore commonly relies on expert parallelism (EP)~\cite{shazeer2017outrageously, jin2026megascale, zhang2025comet, cai2024shortcut},
which partitions experts across GPUs and routes each input token to the GPUs
hosting its selected experts.

EP introduces communication on both sides of every expert
computation. Before the computation, token activations must be \textit{dispatched} to
the GPUs holding their selected experts. The resulting expert outputs must
then be returned to and \textit{combined} on the GPUs that own the tokens. These
operations occur at every MoE layer in a Transformer layer and directly lie on the critical path of inference~\cite{zhang2025comet}. Moreover, recent MoE-based LLMs increasingly employ \textit{fine-grained} experts and activate more experts per token~\cite{krajewski2024scaling}. Although this design improves model capacity and quality without proportionally increasing computation~\cite{liu2024deepseek-v3}, it
increases the number of destinations reached by each token and makes
communication a growing fraction of MoE inference~\cite{guo2026sonicmoe}.

A number of EP communication systems have been developed to
reduce this overhead, including DeepEP~\cite{deepep},
Hybrid-EP~\cite{hybrid-ep}, pplx-kernels~\cite{pplx-kernels}, NCCL
EP~\cite{goldman2026ncclep}, and NIXL~\cite{nixl}. These libraries provide
routing-aware data movement and employ techniques such as device-initiated
communication, fine-grained synchronization, and operation overlapping. 
Their designs, however, assume that GPUs can directly access one
another through NVLink~\cite{nvlink}, GPUDirect Peer-to-Peer (P2P) or RDMA~\cite{gpudirect}. This assumption is fundamental to both their data transfer and synchronization
mechanisms~\cite{ma2026demystifying}.

Such direct GPU access is unavailable on latest consumer GPUs (e.g., RTX 40 and 50 series) which lack NVLink and have PCIe P2P access disabled by the driver. Every
inter-GPU transfer must instead be staged through a bounce buffer in CPU memory~\cite{hu2025demystifying}. Consequently, existing EP communication libraries~\cite{deepep,mao2025uccl,goldman2026ncclep} cannot operate on these systems. Nevertheless, cost-efficient PCIe-based systems equipped with consumer-grade GPUs are widely used for LLM inference by researchers and moderate-scale
institutions~\cite{llamacpp,schultheis2026llmq,borzunov2023petals,
song2024powerinfer,knoop2026private}. To our knowledge, all major LLM frameworks~\cite{kwon2023efficient, zheng2024sglang, liang2025torchtitan, yan2026scalable, rajbhandari2022deepspeed} thus fall back to NCCL~\cite{nccl}, NVIDIA's GPU communication library, that can run without P2P support. Notably, NCCL's newly introduced device API~\cite{nvidia_nccl_228} also requires P2P connectivity and is therefore unavailable on consumer GPUs.

However, the conventional NCCL path available on these systems has several key limitations for MoE inference on consumer GPUs. A traditional ring algorithm~\cite{hammer2025short} implemented in NCCL repeatedly relays data through host memory,
causing the same data to traverse PCIe multiple times. In addtion, NCCL 
communication kernels executed on streaming multiprocessors (SMs) 
compete with expert computation for execution resources~\cite{zhang2025efficient}, limiting their opportunity to overlap with computation~\cite{zhang2025comet}. In addition, inter-GPU synchronization must use completion flags placed in the CPU memory, making every
poll a PCIe transaction. 


In this paper, we present a novel EP communication design for MoE inference on
PCIe-based consumer GPU systems. Our design treats host
memory as a shared communication medium rather than merely as a bounce
buffer between pairs of GPUs. It replaces multi-step data forwarding of conventional ring with single-step collective algorithm that reduce repeated PCIe transfers,
executes bulk data movement on dedicated DMA engines to leave the SMs
available for expert computation overlap, and employs lightweight synchronization
that limits host memory polling and allows each transfer to proceed as soon
as its dependency is satisfied. We integrate the proposed system into vLLM, a state-of-the-art LLM serving engine, and demonstrate that optimizing EP communication effectively translates into end-to-end inference gains on consumer GPU systems.

The key contributions of this paper are as follows:

\begin{itemize}
\item We characterize the inference of MoE models on consumer GPUs and identify the key communication bottlenecks caused by the lack of P2P access, including PCIe link contention, SM competition, and long-latency synchronization through host memory.

\item We propose a novel single-step collective communication algorithm for dispatch and combine operations, which reduceds redundant PCIe transfers incurred when conventional collective algorithms repeatedly relay data via a bounce buffer in CPU memory.

\item We develop an efficient DMA engine-based transfer design and a lightweight synchronization scheme that reduce SM competition, exploit bidirectional PCIe bandwidth, and reduce synchronization overhead.

\item We integrate the proposed design into vLLM and evaluate its performance
at both the collective-operation level and in end-to-end inference on
two PCIe-based systems equipped with NVIDIA RTX 4090 and RTX 5090 GPUs,
respectively.
\end{itemize} 
\section{Background}
\label{sec:background}

In this section, we briefly overview communication patterns in distributed MoE and GPU communication mechanisms.

\subsection{Communication in MoE}

\begin{figure}[t]
\centering
\begin{minipage}{0.70\linewidth}
\begin{subfigure}{0.325\linewidth}%
  \includegraphics[width=\linewidth]{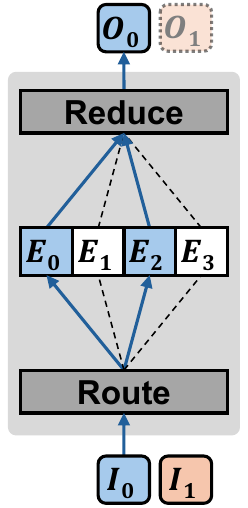}%
  \caption{MoE}%
  \label{fig:single-moe}%
\end{subfigure}%
\hfill%
\begin{subfigure}{0.635\linewidth}%
  \includegraphics[width=\linewidth]{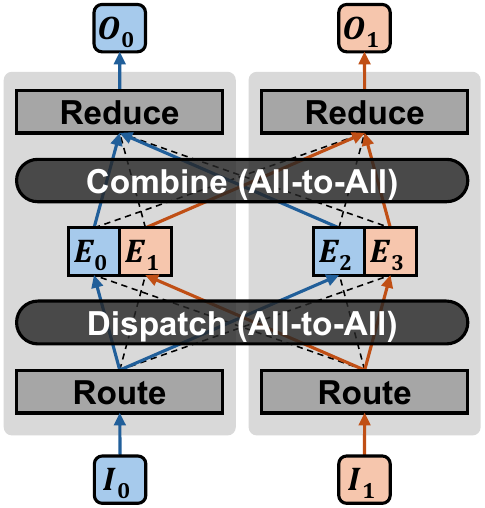}%
  \caption{MoE with EP}%
  \label{fig:distributed-moe}%
\end{subfigure}
\end{minipage}
\caption{A single MoE layer (a) without and (b) with expert parallelism. \textit{I}, \textit{E}, \textit{O} denote input tokens, experts, and outputs.}
\label{fig:comm-in-moe}
\end{figure}

\paragraph{Distributed MoE and expert parallelism} MoE replaces the dense feed-forward network (FFN) of a Transformer layer with $E$ experts and a router.
The router selects $k$ experts per token ($k \ll E$), and only those experts compute on
the token (Figure~\ref{fig:single-moe}). The experts of a modern MoE model typically do not fit in the memory of a single device~\cite{liu2024deepseek-v3}, so they are partitioned across devices under EP~\cite{shazeer2017outrageously}, while attention and the other layers are replicated under data parallelism (DP)~\cite{rajbhandari2020zero}
with each device holding its own tokens. A token and the experts it needs thus sit on
different devices. Each token is sent to the devices holding its selected experts before
the expert computation, and the outputs are returned to the device that owns the token
afterwards. As shown in Figure~\ref{fig:distributed-moe}, MoE on multiple devices with EP, occurs with two All-to-All collective
communications in a single forward pass. The expert computation is thus enclosed by communication on both sides.

\begin{figure}[t]
\centering
\begin{minipage}{0.99\linewidth}
\begin{subfigure}{0.49\linewidth}%
  \includegraphics[width=\linewidth]{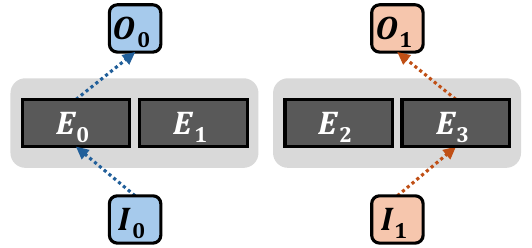}%
  \caption{Coarser (Top-1 on 4 experts)}%
  \label{fig:lower-granularity}%
\end{subfigure}%
\hfill%
\begin{subfigure}{0.49\linewidth}%
  \includegraphics[width=\linewidth]{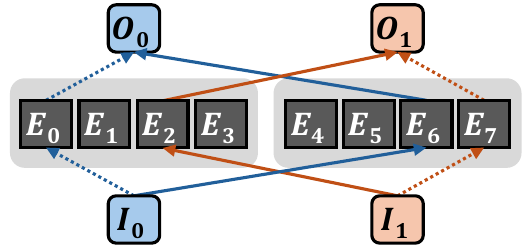}%
  \caption{Finer (Top-2 on 8 experts)}%
  \label{fig:higher-granularity}%
\end{subfigure}
\end{minipage}
\caption{Expert granularity of a MoE layer.}
\label{fig:moe-granularity}
\end{figure}

\paragraph{Recent trend of MoE architectures} Early MoE models such as Mixtral 8x7B~\cite{jiang2024mixtral} have few experts with small top-$k$ routing (8 experts, top-2). Recent models increase the number of experts and use a larger top-$k$ to activate more experts, with each having a smaller hidden dimension to match the same computation cost. DeepSeek-V3~\cite{liu2024deepseek-v3} routes each token to 8 of 256 experts, and Qwen3.8-2.4T-A95B~\cite{qiu2026design} to 10 of 512. 
Figure~\ref{fig:moe-granularity} shows an example of different levels of granularity in
an MoE layer. Finer granularity leaves the per-token computation unchanged. However, communication
does not: each token is copied to $k$ destinations during \textit{dispatch} and gathered back during \textit{combine}, so the transfer volume grows in proportion to $k$ until a token reaches every
remote device, beyond which it cannot grow further. Communication therefore
takes a larger share of the MoE layer as models move toward finer granularity,
which makes optimizing the communication more important than it was
for earlier models.

\paragraph{Communication patterns of dispatch and combine.}

\begin{figure}[t]
\centering
\includegraphics[width=0.70\linewidth]{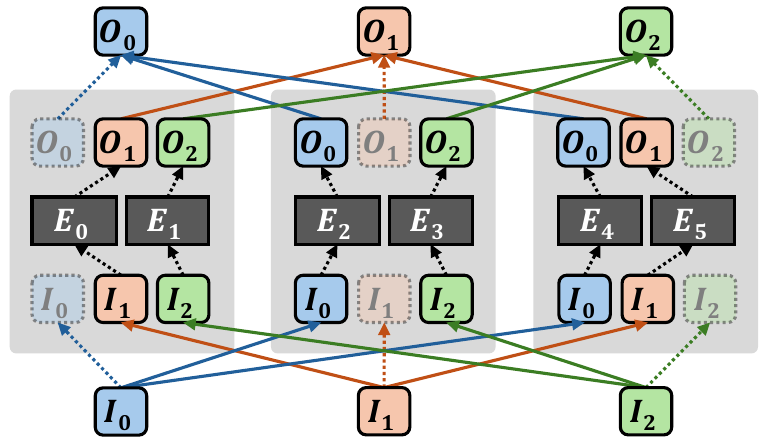}
\caption{An All-Gather and Reduce-Scatter communication pattern for token dispatch and combine in a MoE layer. Top-2 routing with 2 experts on each of 3 devices, where each token selects one expert on each remote device.}
\label{fig:agrs-pattern}
\end{figure}

At the semantic level, token dispatch and combine operations form routing-dependent,
variable-sized many-to-many communication patterns, which can be expressed as \textit{All-to-Allv}~\cite{lei2026fast}. Such an implementation transfers only the tokens and partial
outputs required by the selected experts. An alternative is an
All-Gather and Reduce-Scatter formulation~\cite{jin2026megascale}. During dispatch, each GPU gathers the
input tokens from all other GPUs and locally retains the tokens assigned to its
experts. During combine, each GPU places its partial expert outputs in a dense
buffer ordered by the source tokens, and Reduce-Scatter returns the reduced
outputs to their owning GPUs. Figure~\ref{fig:agrs-pattern} illustrates this
formulation.

The All-Gather and Reduce-Scatter formulation may transfer tokens that are not consumed by any expert
on a receiving GPU. However, it avoids routing-dependent message sizes, CPU-GPU synchronization overhead, and
irregular peer-to-peer exchanges and instead uses regular collective
operations that can achieve higher communication efficiency~\cite{yan2026scalable}. As top-$k$
increases relative to the number of EP devices (i.e., EP size), especially common in recent models, each token tends
to cover more destination GPUs, reducing the redundant communication of the
dense formulation. A recent study similarly reports that All-Gather followed
by Reduce-Scatter outperforms two All-to-All collectives when top-$k$ exceeds the number of
GPUs~\cite{jin2026megascale}. Several state-of-the-art MoE systems, including Megatron-Core and vLLM~\cite{yan2026scalable,kwon2023efficient,zheng2024sglang, tensorrt-llm}, therefore adopt an implementation of All-Gather and Reduce-Scatter as their default EP communication backend.


\subsection{GPU Communication Mechanisms}

\paragraph{Communication between GPUs}

\begin{figure}[t]
\centering
\begin{minipage}{1.0\linewidth}
\begin{subfigure}{0.49\linewidth}%
  \includegraphics[width=\linewidth]{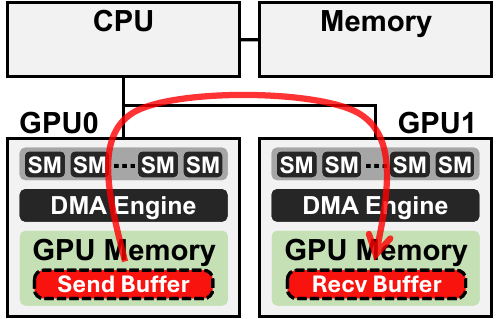}%
  \caption{With P2P}%
  \label{fig:with-p2p}%
\end{subfigure}
\hfill%
\begin{subfigure}{0.49\linewidth}%
  \includegraphics[width=\linewidth]{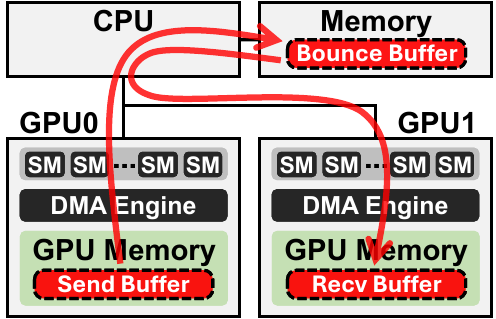}%
  \caption{Without P2P (SHM)}%
  \label{fig:without-p2p}%
\end{subfigure}%
\end{minipage}
\caption{Communication between GPUs in a node.}
\label{fig:p2p}
\end{figure}

GPUs within a same node communicate either directly, namely as \textit{peer-to-peer (P2P)}, or by staging in a bounce buffer in CPU memory, a transport which NCCL calls \textit{shared memory (SHM)}, depending on their interconnect and topology~\cite{hu2025demystifying}. Figure~\ref{fig:with-p2p} shows how intra-node GPUs in a PCIe system communicate with P2P mechanism. For NVIDIA GPUs, P2P is available in two cases. The first is when GPUs are interconnected with a high-bandwidth direct GPU-to-GPU interconnect such as NVLink~\cite{nvlink}. The second is a PCIe-only system in which each device can expose its memory to the PCIe address space so that the other devices can access it directly~\cite{kim2024tccl}. On data-center and workstation GPUs, this direct path, called GPUDirect~\cite{gpudirect} P2P by NVIDIA, is the norm. It is used unless the two GPUs are far apart in the PCIe topology, for example, attached to different CPU root complexes. In that case, NCCL falls back to its SHM transport. Consumer GPUs, however, do not support P2P fundamentally~\cite{schultheis2026llmq}. Therefore, communication between consumer GPUs must pass through the CPU memory. As shown in Figure~\ref{fig:without-p2p}, the sender writes the payload to a pinned host buffer, and the receiver reads it back, so every byte traverses PCIe bus twice~\cite{an2024fire}. 

\paragraph{Data transfer mechanisms}

There are two ways of how GPUs can transfer data each other. First is the DMA or copy engine-based communication. Dedicated DMA engines in the device move data between two memory regions. The transfer is issued as an asynchronous copy such as \texttt{cudaMemcpyAsync}, which is staged through host memory when peer access (i.e., P2P) is unavailable. This mechanism reaches the peak bandwidth of the PCIe link for large messages, but it operates on contiguous or fixed-stride regions specified at issue time and cannot perform data-dependent gather or scatter, so tokens scattered at data-dependent locations must be packed into a contiguous region by a preceding kernel~\cite{nvidia_cuda_bpg_data_transfer}. 

Next, we can use the load/store units (LSUs) in the SMs for GPU-to-GPU communication. Once the destination buffer is mapped into the address space of a device, a GPU kernel reads and writes it with ordinary load and store instructions. NCCL mainly uses this mechanism for some advantages over DMA transfers~\cite{hu2025demystifying}. The transfer granularity is a single access rather than a buffer, which keeps the latency of small messages low and lets a kernel gather or scatter tokens without a separate packing step. The cost is that the transfer occupies SMs that would otherwise run computation~\cite{pati2024t3}.

\section{Related Work}
\label{sec:related}

This section provides related work on communication libraries, DMA engine-based collectives, and communication-centric optimizations for accelerating MoE.

\paragraph{Communication libraries.}
Most GPU communication libraries are designed for systems with direct
GPU-to-GPU access. NCCL~\cite{hu2025demystifying} executes communication
operations using SM kernels and stages data through host memory when P2P
access is unavailable. Its recent device API and DMA engine
path~\cite{nvidia_nccl_228} still require direct P2P connectivity.
Communication libraries specialized for MoE EP, including
DeepEP~\cite{deepep}, pplx-kernels~\cite{pplx-kernels},
NIXL-EP~\cite{nixl}, NCCL EP~\cite{goldman2026ncclep}, and
UCCL-EP~\cite{mao2025uccl}, similarly rely on NVLink, GPUDirect P2P, or
RDMA for their optimized GPU-initiated paths. ThunderEP instead
provides dispatch and combine operations over host memory for consumer GPUs where direct GPU access is unavailable.

\paragraph{Copy engine collectives.}
VCCL~\cite{zhang2025efficient} and FiCCO~\cite{pal2025design} offload
data movement from the SMs to DMA engines, but retain direct
GPU-to-GPU transfers on P2P-capable systems. HFReduce~\cite{an2024fire}
and LLMQ~\cite{schultheis2026llmq} also use DMA engines and host memory,
but target training communication. HFReduce performs CPU-based
All-Reduce for gradients, while LLMQ handles dense weight and gradient
transfers for low-precision pretraining. TCCL~\cite{kim2024tccl}
selects communication paths across PCIe and NUMA topologies, and
llama.cpp~\cite{llamacpp} implements a host-staged two-GPU All-Reduce
for tensor parallelism. ThunderEP targets a different setting by
providing multi-GPU All-Gather and Reduce-Scatter for MoE inference,
retaining reduction on the GPUs and exposing completion by source GPU
to overlap communication with expert computation on consumer GPU
systems without P2P access.

\paragraph{MoE communication optimizations.}
Prior work reduces the cost of EP communication through hierarchical
exchange, as in DeepSpeed-MoE~\cite{rajbhandari2022deepspeed},
Tutel~\cite{hwang2023tutel}, and HetuMoE~\cite{nie2022hetumoe}, or
through topology-aware routing, traffic reduction, and communication
scheduling, as in TA-MoE~\cite{chen2022ta},
ExFlow~\cite{yao2024exploiting}, FasterMoE~\cite{he2022fastermoe},
Lina~\cite{li2023accelerating}, and ScheMoE~\cite{shi2024schemoe}.
Flux~\cite{chang2024flux} and Comet~\cite{zhang2025comet} overlap
communication with computation at a finer granularity, but move data
from GPU kernels over direct P2P links and use SM resources for
communication. Framework techniques such as vLLM's dual-batch
overlap~\cite{vllm_dbo} and SGLang's two-batch
overlap~\cite{sglang_tbo} overlap different microbatches but do not
provide a communication path themselves. For MoE, their communication
overlap relies on device-initiated EP libraries such as
DeepEP~\cite{deepep}, whose optimized paths require NVLink or RDMA.
They therefore do not provide communication overlap on consumer GPU
systems without P2P access. 
ThunderEP addresses the complementary problem of providing an efficient path when communication must pass through CPU memory.

\section{ThunderEP Design}
\label{sec:system}

This section first presents the key problems of optimizing
EP communication for MoE on consumer GPUs and then describes
how the proposed method addresses them.

\begin{figure*}[t]
\centering
\begin{subfigure}{\linewidth}
  \centering
  \includegraphics[width=0.82\linewidth]{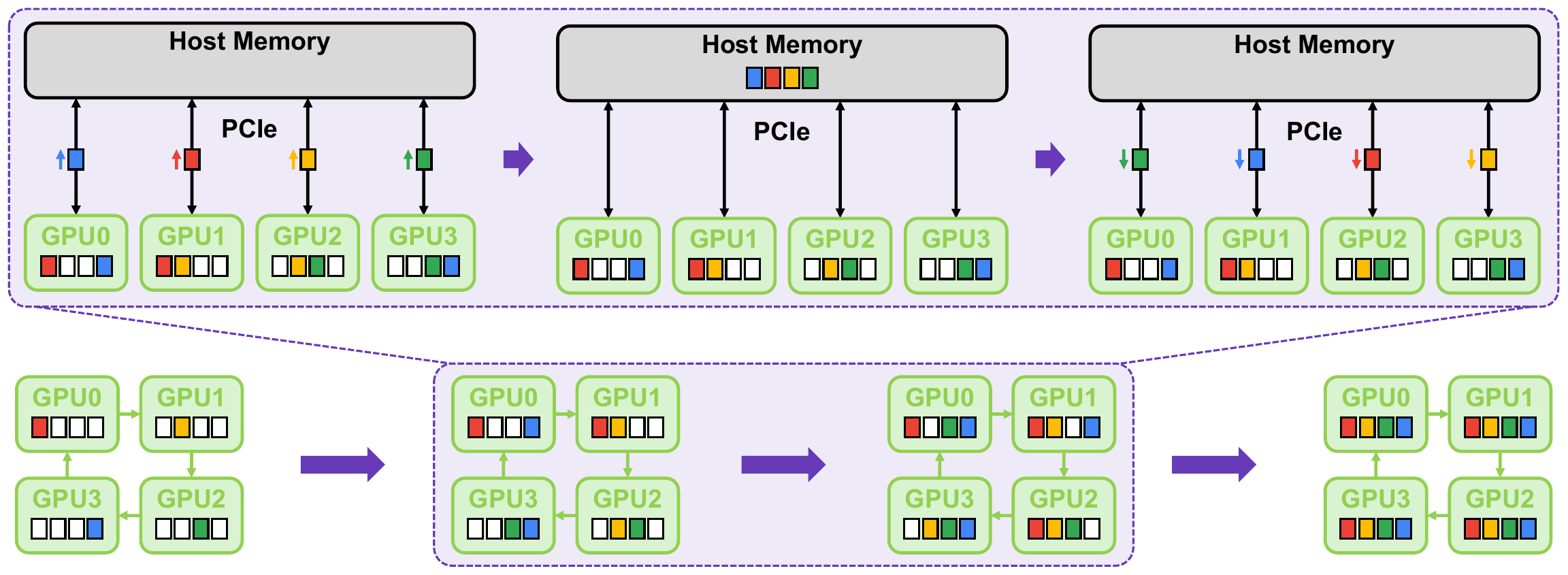}
  \caption{Ring algorithm}
  \label{fig:algo-ag-ring}
\end{subfigure}
\begin{subfigure}{\linewidth}
  \centering
  \includegraphics[width=0.82\linewidth]{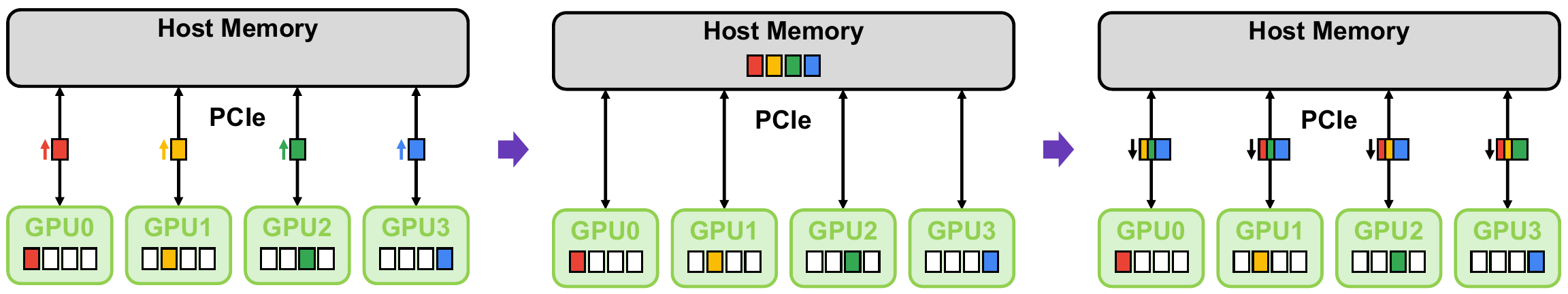}
  \caption{Single-step algorithm}
  \label{fig:algo-ag-ours}
\end{subfigure}
\caption{All-Gather communication algorithm for dispatch in (a) NCCL and (b) ThunderEP.}
\label{fig:algo-ag}
\end{figure*}

\subsection{Challenges}

\paragraph{PCIe link contention}

Without P2P access, communication between consumer GPUs is staged through CPU memory, causing concurrent transfers to contend for shared PCIe link and host memory resources. This contention is particularly costly for MoE EP because all GPUs communicate concurrently during dispatch and combine operations. Measurements in our PCIe-based RTX 5090 system (Table~\ref{tab:system-config}), a GPU communicating alone reaches 56.5~GB/s on PCIe~Gen5, close to its theoretical bandwidth of 64~GB/s. When a second GPU under the same PCIe host bridge communicates concurrently, the bandwidth of each GPU falls to 35.5~GB/s. The bandwidth available to each GPU therefore decreases as more GPUs participate in a collective, making it important to minimize both the PCIe traffic volume and repeated transfers through host memory.

\paragraph{SM competition}

NCCL performs its collective operations using kernels that occupy the GPU SMs. When communication overlaps with expert computation, the two operations compete for the same SM resources~\cite{schultheis2026llmq, zhang2025efficient}. We observe that co-scheduling an NCCL with a GEMM delays the communication kernel and stalls the progress. DMA engines avoid this contention by moving data without occupying the SMs, but each transfer incurs a fixed issue overhead and can access only contiguous or fixed stride regions. Data stored at routing-dependent locations must therefore be packed before transfer, and the issue overhead is particularly significant for the small transfers encountered during decoding. Although recent NCCL releases provide collective communication using DMA engines~\cite{nvidia_nccl_228}, this path requires P2P access and cannot operate on consumer GPU systems.

\paragraph{Long-latency synchronization}

A collective communication further depends on synchronization among the participating GPUs, since a receiver must learn when a sender has finished writing the data destined for it. NCCL signals completion through a counter or flag buffer, and the collective kernel itself polls this buffer, occupying an SM and busy-waiting until the expected value appears.
On consumer GPUs, where no GPU can address another GPU's memory, the only memory both GPUs
can map is the CPU memory. The completion flag buffer must therefore reside in pinned host
memory, and every check of it is a round trip across the PCIe link, made by a kernel
that remains resident on an SM.

\paragraph{ThunderEP's approach.}
ThunderEP aims to enable efficient communication-computation overlap
for MoE EP on consumer GPUs without P2P support. Our communication
optimizations and collective design follow three principles. First, because concurrent GPU transfers contend
for shared PCIe resources, it is important to minimize both the communication
volume and the number of times data traverses PCIe. Second, because
SM-resident communication competes with expert computation, communication
should use as few SM resources as possible to enable effective overlap with
expert GEMMs. Third, because inter-GPU synchronization relies on flags placed
in host memory, it is essential to minimize expensive polling traffic and
unnecessary waiting among GPUs. Based on these requirements, we design a new algorithm (§~\ref{sec:algorithm}), DMA engine-based communication design (§~\ref{sec:dma}), and efficient synchronization scheme (§~\ref{sec:sync}) for
MoE communication on consumer GPUs.

\begin{figure*}[t]
\centering
\begin{subfigure}{\linewidth}
  \centering
    \includegraphics[width=0.82\linewidth]{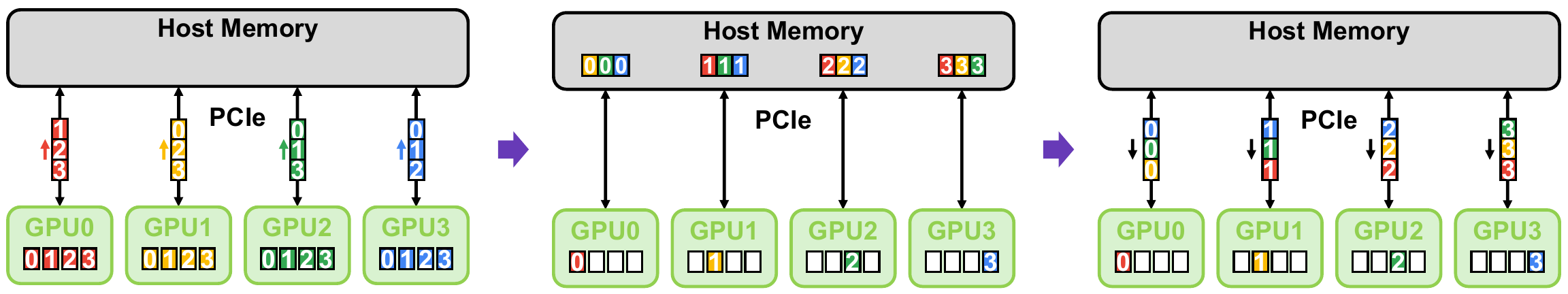}
  \caption{GPU reduction}
  \label{fig:algo-rs-gpu}
\end{subfigure}
\begin{subfigure}{\linewidth}
  \centering
  \includegraphics[width=0.82\linewidth]{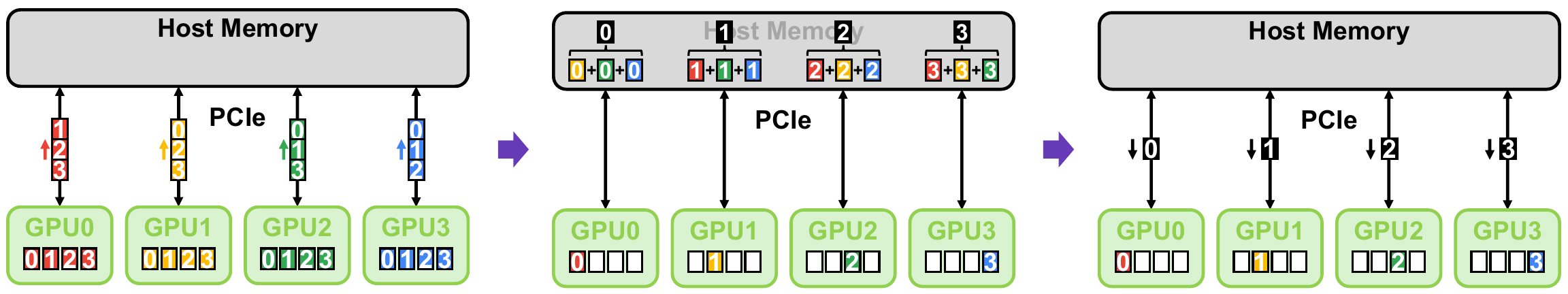}
  \caption{Partial CPU reduction}
  \label{fig:algo-rs-cpu}
\end{subfigure}
\caption{Proposed Reduce-Scatter algorithm for combine with (a) GPU reduction and (b) CPU reduction.}
\label{fig:algo-rs}
\end{figure*}

\subsection{Single-step Collective Algorithm}
\label{sec:algorithm}

NCCL performs most of its collective communication operations primarily with a ring algorithm~\cite{hu2025demystifying}. The traditional ring algorithm is a bandwidth-optimal, circular scheme in which each of the $N$ GPUs in a node, where $N$ is the number of GPUs, forwards the \textit{shard} (i.e., one GPU's data partition) it has just received on to its successor, so that after $N-1$ steps every GPU holds the data of all $N$ participants. On a system with NVLink or GPUDirect P2P mechanism available, this design reaches close to the system's peak bandwidth, because every link carries only the $(N-1)/N$ fraction of the data that it must and each forwarding step is a direct GPU-to-GPU copy~\cite{hu2025demystifying}.

On a PCIe system, where the GPUs are commodity cards (e.g., NVIDIA Geforce RTX series) without P2P support, this reasoning no longer holds. Every transfer is staged through CPU memory, so a single step is not a direct copy but a round trip that writes the shard into host memory and then reads it back out on the next GPU, causing redundant data transfer as shown in Figure~\ref{fig:algo-ag-ring}. We further find that the effective PCIe bandwidth does not stay constant as more GPUs communicate at once, since the transfers contend for the shared PCIe link (e.g., host bridge) and the host memory behind it. Under this contention, spreading a collective across many links buys little, and it is more profitable to shorten the number of hops that each byte takes and to reduce the total volume that crosses the PCIe links in the first place.

To reduce the redundant transfer, we therefore replace the multi-step ring's relay with a single-step broadcast via a bounce buffer on the CPU memory in All-Gather (i.e., dispatch), shown in Figure~\ref{fig:algo-ag-ours}. First, each GPU writes its own shard once into host memory buffer, and then, every GPU reads the remaining $N-1$ shards back directly. No shard is ever forwarded between GPUs, so the pinned host memory serves as a single broadcast medium that each GPU writes to once and
reads from as needed. The two schedules (i.e., NCCL's ring and the single-step algorithm) differ in both the volume they move and the number of hops each byte travels. Let $S$ denote the size of one GPU's shard. The ring transfers $2(N-1)S$ bytes per GPU, one upload and one download at each of its $N-1$ sequential steps, and a shard is relayed through as many as $N-1$ host round trips before it reaches the farthest GPU. ThunderEP transfers $N \cdot S$
bytes per GPU, a single upload of its own shard and $N-1$ downloads of the others, and it completes in a fixed two phases regardless of how many GPUs participate. The ratio of PCIe transfer volume between
the two is $2(N-1)/N$, which equals one at $N=2$ and grows toward two as the node fills. Our design therefore breaks even at two GPUs compared to NCCL ring and begins to save once the node holds more than two, and the advantage widens as the GPU count grows.

Reduce-Scatter for combine operation follows the exact reverse data flow of
All-Gather for dispatch, but must additionally sum the \textit{blocks} (i.e., each GPU's contribution to an output shard) produced by different GPUs. Figure~\ref{fig:algo-rs} shows two
choices for placing this reduction. With GPU reduction
(Figure~\ref{fig:algo-rs-gpu}), each GPU uploads the $N-1$ blocks destined for other GPUs to CPU memory, and each destination GPU
downloads the $N-1$ remote blocks intended for it and sums them
with its local block to form its output shard. This design
eliminates the sequential relays and synchronization of the ring, but
moves $2(N-1)S$ bytes per GPU, the same analytic communication volume
as NCCL. With partial CPU reduction
(Figure~\ref{fig:algo-rs-cpu}), the CPU first sums the remote GPUs' blocks for each destination in host memory, so that the destination GPU
downloads a single reduced result and adds its local contribution. This design uploads $(N-1)S$ bytes and downloads $S$ bytes per
GPU, resulting in a total volume of $N \cdot S$, the same as ThunderEP's
single-step algorithm for All-Gather with the transfer directions
reversed.

\begin{figure}[t]
\centering
  \includegraphics[width=0.80\linewidth]{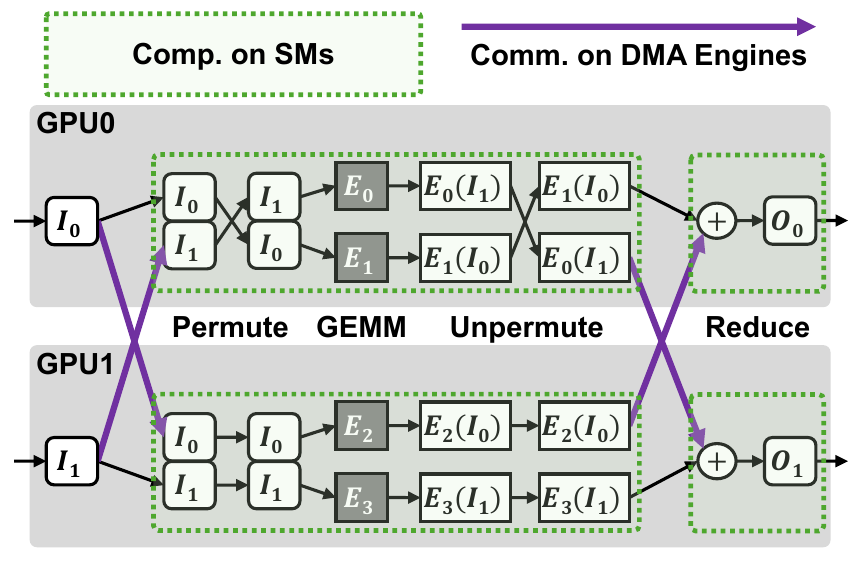}%
\caption{DMA engine-based communication for token dispatch and combine in MoE with EP.}
\label{fig:dma-engine}
\end{figure}

\subsection{DMA Engine-based Communication}
\label{sec:dma}

NCCL drives the data transfer with SM-resident kernels that copy each shard through a bounce buffer in the CPU memory on a consumer-grade GPU node. This approach takes SMs away from computation, and the communication kernels must first be launched and scheduled onto those SMs, which adds latency and, when the SMs are occupied by the expert GEMM, stalls the transfer until they free up~\cite{rashidi2021enabling, zhang2025efficient}. Therefore, we instead issue transfers on the GPU's DMA engines, the dedicated hardware that crosses PCIe with no SM and without waiting for other computation kernels.

Figure~\ref{fig:dma-engine} describes a single MoE layer under this design. The dispatch and combine communication cross PCIe on the DMA engines, while the expert
computation (i.e., grouped GEMM) between them runs on the SMs. On dispatch, each token must
reach the GPUs that hold its experts. Our All-Gather realizes this by copying every GPU's tokens
to every other GPU on the DMA engine. Each GPU then groups the tokens by the experts it owns,
runs its expert GEMMs, and returns the outputs to token order, a permutation that every
grouped MoE performs and that stays local on the SMs. On combine, the DMA engine copies
each GPU's expert outputs back toward the tokens' owners, and an SM adds the contributions that
arrive for each token. This reduction must stay on an SM, since the DMA engine can only move
bytes and has no ALUs to sum them~\cite{schultheis2026llmq}. Because the DMA engine carries both
transfers as large contiguous regions~\cite{nvidia-cuda-bpg}, the SMs remain
free to run the expert GEMMs while the bytes are in flight, and an SM is spent only on the
combine's reduction.

\begin{figure}[t]
\centering
\begin{subfigure}{\linewidth}
  \centering
  \includegraphics[width=0.80\linewidth]{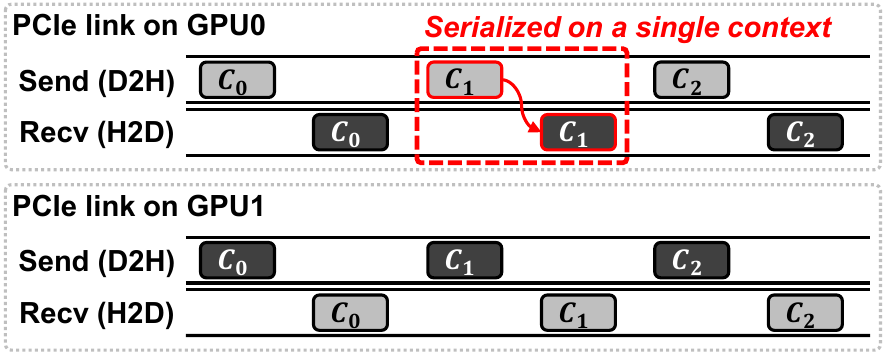}
  \caption{Serialized (single stream)}
  \label{fig:stream-single}
\end{subfigure}
\begin{subfigure}{\linewidth}
  \centering
  \includegraphics[width=0.80\linewidth]{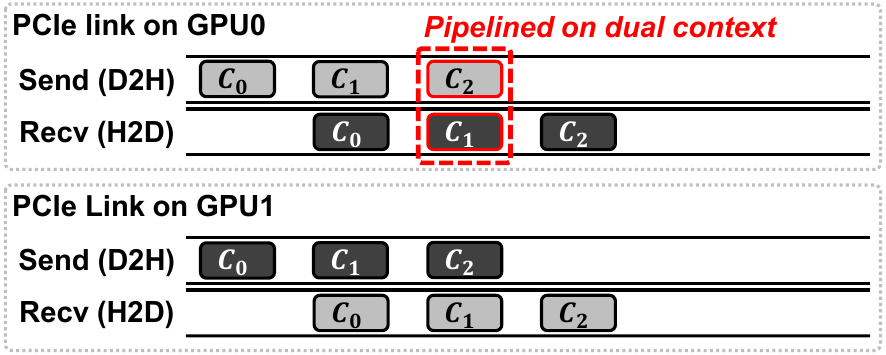}
  \caption{Pipelined (dual stream)}
  \label{fig:stream-dual}
\end{subfigure}
\caption{(a) Serialized and (b) pipelined  bidirectional chunked data transfer over a full-duplex PCIe link.}
\label{fig:duplex-chunk}
\end{figure}

In these communications, every GPU sends and receives data at the same time, so both directions of its PCIe link are active together. A PCIe link is full-duplex, so it can carry a send on the upstream
lane and a receive on the downstream lane at once. We exploit this by placing outgoing DMA transfers on one stream
and incoming transfers on a second, and by splitting each transfer into chunks so that the
upload of one chunk overlaps the download of the previous one, as illustrated in
Figure~\ref{fig:stream-dual}. The overlap comes from giving each direction its
own execution context, an in-order stream or kernel, since a single context
serializes the two directions whether it carries them on a DMA stream or an SM
kernel, as in the single stream case of Figure~\ref{fig:stream-single}. The DMA
engine then earns its place in two ways. It runs the two directions with no SM
cost, leaving the processing units free for the expert GEMM, and a single DMA
queue is deep enough to saturate its direction on its own, so two streams are
enough.

However, NCCL does not reach this overlap on our fabric. It drives a collective from a GPU kernel using SMs, and a barrier within each communication channel forces the send of an internal chunk to complete before its receive begins. The two directions thus alternate chunk by chunk, while multiple channels increase bandwidth rather than overlap because each remains subject to the same barrier~\cite{hu2025demystifying}. The NCCL collective therefore behaves like the serialized single stream transfer
of Figure~\ref{fig:stream-single}. This behavior is specific to systems without P2P. With P2P a GPU's peers write their chunks straight into its memory, so its own kernel only
sends while the receiving takes care of itself, and the two directions overlap.
Without P2P every chunk goes through host memory, so the kernel must both write
its own out and read the incoming ones back in, and NCCL's warp-group barrier orders the two phases rather than issuing them together.

The number of chunks has a trade-off. Too few and the pipeline
spends most of its time filling and draining with one lane idle at each end, too
many and the per-chunk handoff between the streams costs more than the overlap
it buys, so we use the smallest chunk count that keeps both lanes busy through
the steady state.


\begin{figure}[t]
\centering
\begin{subfigure}[b]{0.720\linewidth}
\centering
\begin{minipage}{\linewidth}
  \includegraphics[width=\linewidth]{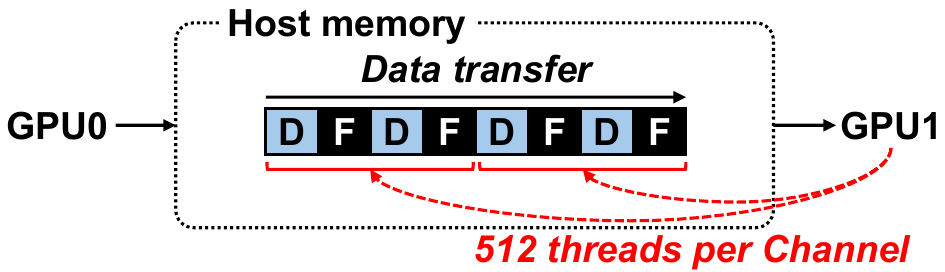}
\end{minipage}
\caption{NCCL's LL protocol}
\label{fig:sync-nccl}
\end{subfigure}
\hfill
\begin{subfigure}[b]{0.720\linewidth}
\centering
\begin{minipage}{\linewidth}
  \includegraphics[width=\linewidth]{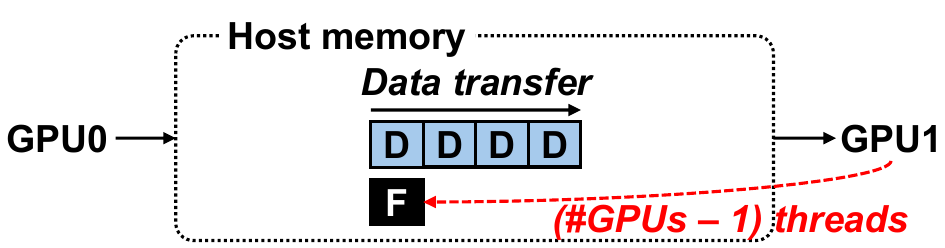}
\end{minipage}
\caption{Proposed method}
\label{fig:sync-ours}
\end{subfigure}
\caption{Comparison of the synchronization protocols used by NCCL and the proposed method.
(a) NCCL pairs every 4 bytes of data with a 4 byte completion flag.
(b) ThunderEP stores completion flags separately from the data and uses a small number of threads to poll them.
$D$ denotes data, and $F$ denotes a completion flag.}
\label{fig:sync-kernel}
\end{figure}

\subsection{Efficient Synchronization Strategy}
\label{sec:sync}

A GPU can read transferred data only after the sender finishes writing
it, so every transfer requires a completion signal. Without P2P access,
host memory is the only memory region accessible to both GPUs, and the
signal must be stored there. Reading this signal requires a PCIe round
trip. For large message transfers in prefill, this cost is hidden by the data
transfer. However, for small transfers in decode, it can dominate the communication latency.

Figure~\ref{fig:sync-kernel} compares the completion signaling used by
NCCL and ThunderEP. NCCL's LL (Low Latency) protocol pairs every 4 bytes of data with
a 4 byte flag, as shown in Figure~\ref{fig:sync-nccl}. An SM writes the
data and its associated flag within a single instruction, so observing
the flag also confirms that the data has arrived. This design is
efficient with P2P access, where the data and flags reside in GPU
memory. On consumer GPUs, NCCL stages both in host memory, turning every
flag check into a PCIe round trip. The inline flags also double the
space required for a given payload and split the data into small
pieces. In addtion, NCCL's LL uses up to 512 threads per channel to check these flags, and an unsuccessful check still fetches the data associated with the
flag.

The LL protocol's inline flag format is not directly compatible with DMA-based transfers. A DMA
engine copies a source buffer as laid out and cannot insert flag words
between payload elements. Constructing the interleaved layout in
advance would require an additional SM packing operation, defeating the
purpose of using DMA to avoid SM contention. Even with such a layout, a
DMA engine does not guarantee the arrival order of individual lines, so
observing a flag would not ensure that its associated data had arrived.
ThunderEP therefore keeps the payload contiguous and stores completion
flags separately, as shown in Figure~\ref{fig:sync-ours}. One thread per
remote sender polls its flag with uncached loads and pauses between
attempts, while the remaining threads wait on a word in GPU memory.
Each flag has a single writer, so no atomic operation is required. The
main benefit comes not from a faster individual flag check, but from
avoiding repeated checks across the $N-1$ hops of the ring through
ThunderEP's single-step algorithm.

\begin{figure}[t]
\centering
\begin{subfigure}[b]{\linewidth}
\centering
\begin{minipage}{0.88\linewidth}
  \includegraphics[width=\linewidth]{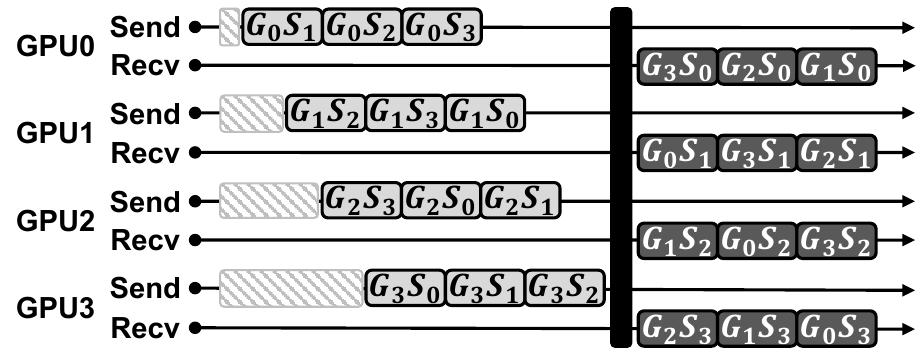}
\end{minipage}
\caption{Waiting for all senders}
\label{fig:sync-global}
\end{subfigure}
\vspace{0.6em}
\begin{subfigure}[b]{\linewidth}
\centering
\begin{minipage}{0.88\linewidth}
  \includegraphics[width=\linewidth]{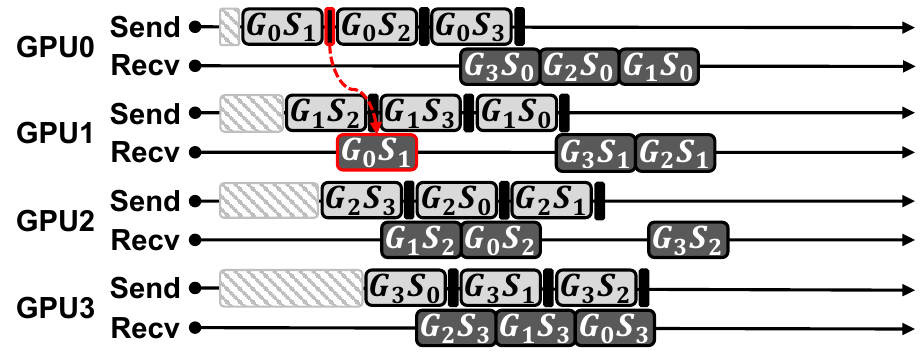}
\end{minipage}
\caption{Waiting for a single sender}
\label{fig:sync-peer}
\end{subfigure}
\caption{Fine-grained synchronization policy for the Reduce-Scatter in combine, which
  mitigates the idle time when senders finish at different times because the
  router gives each device a different amount of expert work. 
  (a) No receive begins before the last sender finishes.
  (b) Each begins on its own sender's completion flag.
  $G_pS_d$ is GPU $p$'s contribution block for shard $d$, reduced on GPU $d$.}
\label{fig:sync-policy}
\end{figure}

ThunderEP also avoids waiting for all senders before receiving.
Figure~\ref{fig:sync-policy} illustrates this design, using combine as
an example. Routing assigns different amounts of expert work to the
GPUs, and PCIe contention can further increase the difference in their
completion times. With the global policy in
Figure~\ref{fig:sync-global}, every GPU waits for the slowest sender.
With the per-sender policy in Figure~\ref{fig:sync-peer}, a GPU begins
reading from each sender as soon as its completion flag is available.
During combine, a sender also releases each destination's portion as
soon as it is written rather than waiting for its entire output.
Transfers from ready senders can therefore proceed while slower senders
are still working.

Removing such global synchronization requires preventing a fast sender from
overwriting a host buffer that a slower receiver is still reading.
ThunderEP alternates between two buffers and reuses a buffer only after
the receivers report that they have consumed its previous contents.
Receivers send these credits through PCIe writes. Writes do not wait
for a response, whereas reads do and saturate earlier, so this
notification avoids the more constrained read path. Because each
buffer is reused only every other round, the credits normally arrive
before it is needed again. This prevents overwrites without delaying
the next upload.

These techniques benefit different phases. Separating the flags and
limiting the polling threads primarily reduce decode latency because
flag checks account for a large fraction of small transfer time. In
prefill, this overhead is hidden by the larger data transfer. Per-sender
waiting policy is more useful in prefill because the larger expert workload
creates enough difference in completion time to overlap receiving with
the remaining sends. Decode transfers are too small for this overlap
to provide a comparable benefit.





\subsection{Integration and Implementation}
\label{sec:implementation}

We integrate ThunderEP into vLLM~\cite{kwon2023efficient}, a
state-of-the-art LLM inference engine, by replacing its NCCL-based
All-Gather and Reduce-Scatter operations for EP dispatch and combine in
both the prefill and decode phases. During decode, vLLM replays the
execution with a CUDA graph to reduce launch overhead. ThunderEP
alternates between two staging buffers in host memory so that the next
communication round can begin while receivers finish reading the
previous round, avoiding an additional synchronization between rounds. A
captured \texttt{cudaMemcpyAsync} operation for a DMA transfer reuses
the address recorded at capture time, so alternating buffers would
require the host to update the graph before every replay. ThunderEP
therefore performs the data transfers of the decode phase on the SMs,
where the kernel itself reads the round number from device memory and
picks the buffer, leaving the captured graph untouched. Little is lost
by doing so, because at decoding batch sizes both the transfers and the
expert GEMM are small and the SMs are not a contended resource. The
router, expert placement, and computation kernels remain unchanged.

\paragraph{Communication-computation overlap.}
A prefill step carries enough tokens to pipeline communication and
expert computation. Decode transfers only a few kilobytes per GPU and
provides too little work for such a pipeline, so ThunderEP applies
overlap only during prefill. The host submits transfers and expert
computation on each GPU's token activations to separate CUDA streams and
connects their dependencies with CUDA events. Computation begins on each
GPU's activations as soon as they arrive, while transfers from the
remaining GPUs continue. The resulting outputs are returned to that GPU
as soon as the computation finishes. Existing device-initiated EP
libraries also support communication and computation overlap through
hooks or separate send and completion
interfaces~\cite{deepep,mao2025uccl}, but they require peer memory that
device code can address or a NIC the GPU can drive. In addition, the
caller observes completion only for an entire dispatch or combine
operation, even when the communication kernel tracks individual peer
arrivals internally. ThunderEP instead uses the fine-grained synchronization described in
Section~\ref{sec:sync} to expose completion per sender. Computation can
therefore begin upon each sender's completion, without waiting for all
senders.

\paragraph{Reduction placement}
ThunderEP implements both GPU and partial CPU reduction for
Reduce-Scatter, as shown in Figure~\ref{fig:algo-rs}. Partial CPU
reduction combines remote partial results in host memory, allowing each
GPU to download only a single reduced result and thereby reducing
PCIe traffic. However, this reduction accesses the same host memory that
serves concurrent GPU DMA transfers. On our systems, the aggregate PCIe
bandwidth exceeds the available host DRAM bandwidth, leaving little
bandwidth for CPU reduction. The resulting CPU reduction overhead exceeds
the time saved by reducing PCIe traffic. We therefore adopt GPU reduction
in ThunderEP.

\paragraph{Routing-aware communication} 
We also implement routing-aware dispatch and combine operation that transfer only the data required by each destination. On consumer GPUs, however, DMA transfers are issued through host APIs, while routing decisions are generated on the GPU. The CPU must therefore retrieve the routing metadata before issuing transfers at every MoE layer, introducing CPU-GPU synchronization~\cite{garg2026moefusion} and numerous small DMA requests. In our measurements, these overheads offset the reduction in PCIe traffic. GPU-initiated EP libraries~\cite{deepep,goldman2026ncclep,hamidouche2025gpu} avoid host coordination through NVLink or RDMA, which our target systems do not support. During decode, a fixed CUDA graph cannot replay
routing-dependent DMA descriptors without host updates. ThunderEP therefore retains All-Gather and Reduce-Scatter and leaves efficient routing-aware All-to-All communication on consumer GPUs to future work.

\begin{figure*}[t]
  \centering
  \begin{subfigure}[t]{0.245\textwidth}
    \centering
    \includegraphics[width=\linewidth]{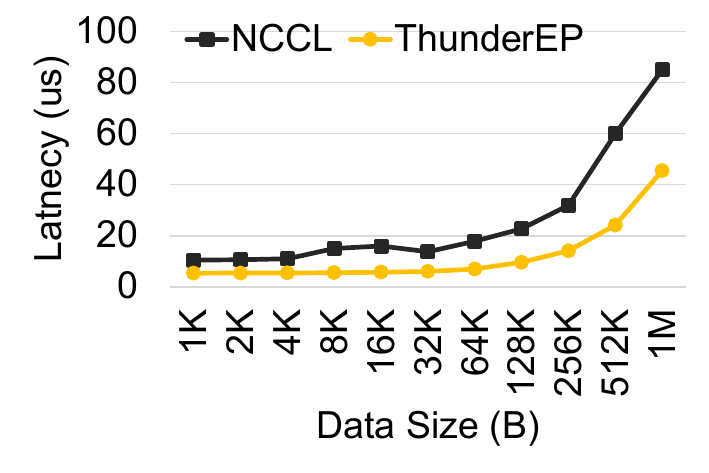}
    \caption{RTX~4090 (1KB--1MB).}
    \label{subfig:ag-lat-4090}
  \end{subfigure}
  \hfill
  \begin{subfigure}[t]{0.245\textwidth}
    \centering
    \includegraphics[width=\linewidth]{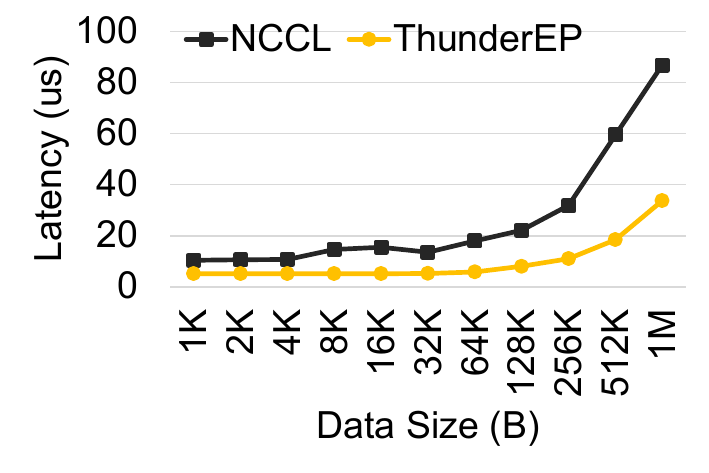}
    \caption{RTX~5090 (1KB--1MB).}
    \label{subfig:ag-lat-5090}
  \end{subfigure}
  \hfill
  \begin{subfigure}[t]{0.245\textwidth}
    \centering
    \includegraphics[width=\linewidth]{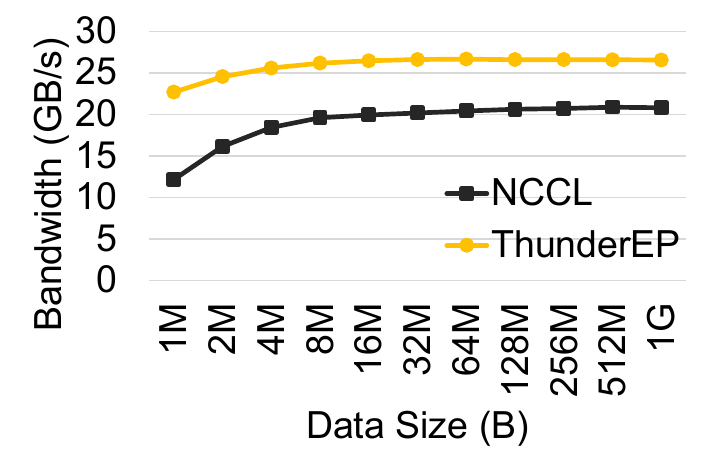}
    \caption{RTX~4090 (1MB--1GB).}
    \label{subfig:ag-bw-4090}
  \end{subfigure}
  \hfill
  \begin{subfigure}[t]{0.245\textwidth}
    \centering
    \includegraphics[width=\linewidth]{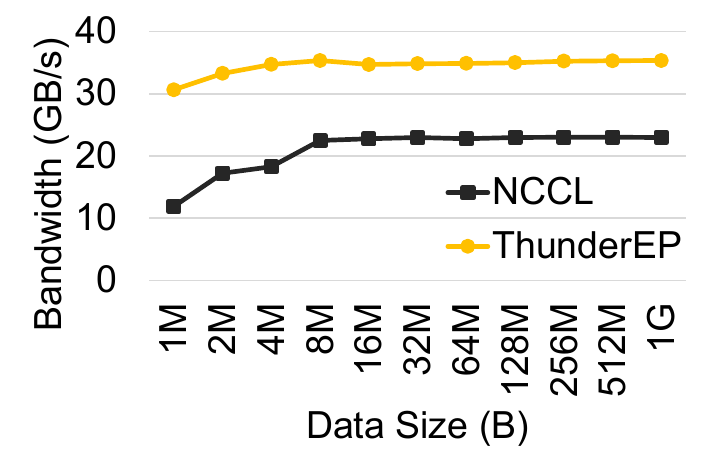}
    \caption{RTX~5090 (1MB--1GB).}
    \label{subfig:ag-bw-5090}
  \end{subfigure}
  \caption{All-Gather performance for token dispatch on a single node (6 GPUs).}
  \label{fig:ag-perf}
\end{figure*}

\begin{figure*}[t]
  \centering
  \begin{subfigure}[t]{0.245\textwidth}
    \centering
    \includegraphics[width=\linewidth]{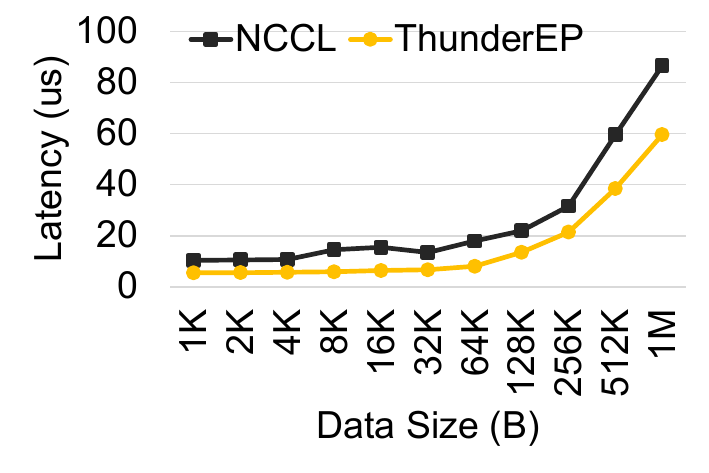}
    \caption{RTX~4090 (1KB--1MB).}
    \label{subfig:rs-lat-4090}
  \end{subfigure}
  \hfill
  \begin{subfigure}[t]{0.245\textwidth}
    \centering
    \includegraphics[width=\linewidth]{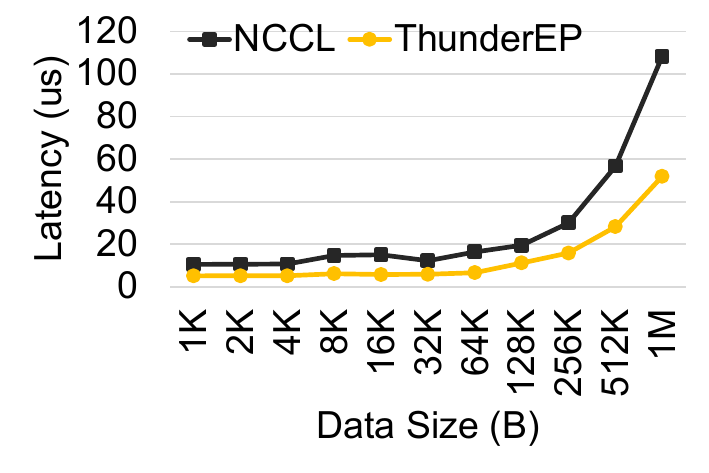}
    \caption{RTX~5090 (1KB--1MB).}
     \label{subfig:rs-lat-5090}
  \end{subfigure}
  \hfill
  \begin{subfigure}[t]{0.245\textwidth}
    \centering
    \includegraphics[width=\linewidth]{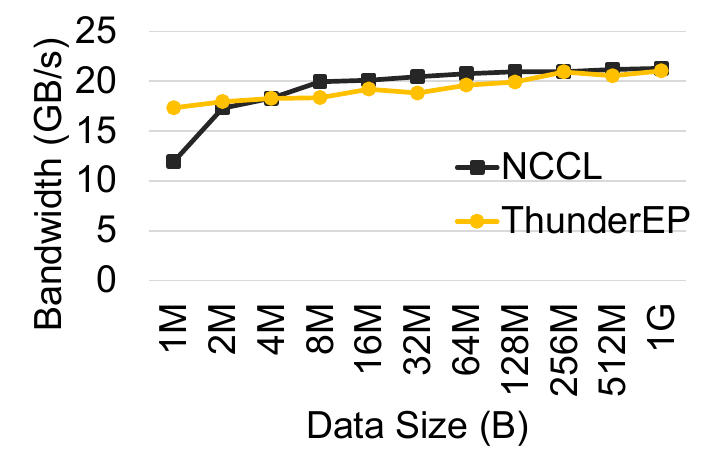}
    \caption{RTX~4090 (1MB--1GB).}
     \label{subfig:rs-bw-4090}
  \end{subfigure}
  \hfill
  \begin{subfigure}[t]{0.245\textwidth}
    \centering
    \includegraphics[width=\linewidth]{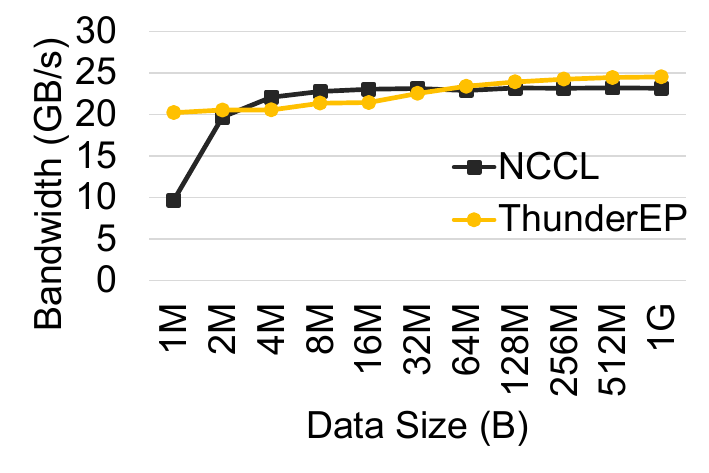}
    \caption{RTX~5090 (1MB--1GB).}
    \label{subfig:rs-bw-5090}
  \end{subfigure}
  \caption{Reduce-Scatter performance for token combine on a single node (6 GPUs).}
  \label{fig:rs-perf}
\end{figure*}

\section{Evaluation}
\label{sec:eval}

We evaluate the communication and end-to-end performance
of the proposed method on two PCIe-only consumer GPU systems.
We first evaluate the communication performance of dispatch and combine, 
and then measure their impact on end-to-end LLM inference.

\subsection{Experimental Environment}

\paragraph{System configurations}
We conduct the experiments on two single-node PCIe systems equipped with
6 $\times$ NVIDIA GeForce RTX 4090 and RTX 5090 GPUs, respectively.
In each system, the GPUs are distributed across three PCIe host
bridges, with two GPUs attached to each host bridge, while all devices
belong to a single NUMA node.
Neither system supports GPUDirect P2P access between any pair of GPUs.
Consequently, communication between GPUs must be staged through host
memory.
Table~\ref{tab:system-config} summarizes the details of our system configurations.


\begin{table}[t]
  \centering
  \small
  \caption{System configurations used in our evaluation.}
  \label{tab:system-config}
  \vspace{-0.5em}
  \begin{tabular}{|l|c|c|}
    \hline                  
                        Systems
                        & RTX~4090 system
                        & RTX~5090 system \\
    \hline\hline
    Mainboard
                        & \multicolumn{2}{c|}{ASRock Rack GENOAD8X-2T} \\
    \hline
    CPU
                        & \multicolumn{2}{c|}{1 $\times$ AMD EPYC~9124
                          (16-Core)} \\
    \hline
    Main memory
                        & \multicolumn{2}{c|}{8 $\times$ DDR5-4800 32\,GB} \\
    \hline
    GPU
                        & 6 $\times$ RTX~4090
                        & 6 $\times$ RTX~5090 \\
    \hline
    GPU memory
                        & 24\,GB
                        & 32\,GB \\
    \hline
    GPU interconnect
                        & PCIe~4.0 $\times$16
                        & PCIe~5.0 $\times$16 \\
    \hline
    OS / Kernel
                        & \multicolumn{2}{c|}{Ubuntu~24.04.3 LTS
                          / Linux~6.8.0-35} \\
    \hline
    GPU driver
                        & \multicolumn{2}{c|}{580.65.06 (CUDA 13.0)} \\
    \hline
  \end{tabular}
\end{table}

\paragraph{Comparison baselines.}
We use vLLM (v0.27.1)~\cite{kwon2023efficient} as our primary baseline to evaluate the performance benefit of replacing its default EP communication backend with ThunderEP. vLLM performs MoE dispatch and combine using NCCL All-Gather and Reduce-Scatter, respectively.
ThunderEP replaces only this communication path and all other computation kernels and execution settings remain unchanged.

We additionally compare end-to-end inference performance against SGLang (v0.5.18)~\cite{zheng2024sglang} and Megatron-Core (v0.19.2)~\cite{yan2026scalable}. SGLang and Megatron-Core also employ an All-Gather and Reduce-Scatter communication pattern for EP rather than All-to-Allv. For each experiment, all systems use the same numerical precision for computation and communication with identical EP and DP configurations.

To our knowledge, existing EP-specialized communication libraries~\cite{deepep, hybrid-ep, pplx-kernels, goldman2026ncclep, nixl} cannot run on our target systems because they rely on NVLink, GPUDirect P2P, or GPUDirect RDMA. SGLang and Megatron-Core therefore also use NCCL backend for EP communication on our systems.
For collective operation-level comparisons, we compare ThunderEP directly against the corresponding NCCL (v2.30.7)~\cite{nccl} primitives using identical message sizes and GPU counts.

\paragraph{Models and Workloads.}
We evaluate three popular open-weight MoE-based LLMs:
Qwen3-30B-A3B~\cite{yang2025qwen3} and two GPT-OSS models,
GPT-OSS-20B and GPT-OSS-120B~\cite{agarwal2025gpt}. We use BF16 for Qwen3-30B-A3B, and use the native MXFP4 checkpoint for GPT-OSS models.

For workloads, we use both synthetic and real-world datasets. We primarily use synthetic prompts to vary prompt lengths and batch sizes, allowing us to comprehensively evaluate prefill and decode across a wide range of workload sizes. To verify that the observed performance trends are not specific to synthetic workloads, we additionally employ prompts sampled from ShareGPT~\cite{sharegpt} and LMSYS-Chat-1M~\cite{zheng2024lmsys}, two widely used real-world LLM conversation datasets. All results report the average of five runs.

\subsection{Communication Performance}

We report messages from 1KB to 1MB in terms of latency, where startup and synchronization overheads dominate, and messages from 1MB to 1GB in terms of bandwidth, where data movement dominates~\cite{hwang2026msccl++, kim2026hetccl}.
The two ranges represent the communication behavior of decode and prefill, respectively.
Message size denotes the full array size for each collective operation. We observe consistent performance trends on both the RTX 4090 and RTX 5090 systems.

\paragraph{Dispatch.}
Figure~\ref{fig:ag-perf} compares the All-Gather performance for token dispatch.
Across the 1KB--1MB range, ThunderEP achieves average latency speedups of 2.30$\times$ and 2.65$\times$ on the RTX~4090 and RTX~5090 systems, respectively.
The main benefit comes from eliminating the sequential relays of NCCL's ring algorithm, thereby reducing both PCIe traffic and synchronization latency through pinned host memory.
With total 6 GPUs and a per-GPU shard of size $S$, ThunderEP reduces the analytic PCIe traffic from $10S$ to $6S$, replaces the five steps of ring path with a direct host-staged transfer, and reduces the number of synchronization steps from five with NCCL's LL protocol to two. The reduced synchronization count dominates latency for small messages, whereas the lower traffic volume becomes the primary factor for large messages.

The bandwidth advantage persists throughout the 1MB--1GB range. At 1GB, where both implementations have reached their sustained bandwidth, ThunderEP reaches 26.5 GB/s and 35.3 GB/s on the RTX~4090 and RTX~5090 systems, respectively, compared with 20.8 GB/s and 23.0 GB/s for NCCL. In addition to moving 40\% fewer bytes, ThunderEP overlaps D2H and H2D transfers on separate DMA streams and uses both directions of the full-duplex PCIe link.
NCCL instead serializes the two directions within each ring channel. 


\begin{figure*}[t]
  \centering
  \begin{subfigure}[t]{0.32\textwidth}
    \centering
    \includegraphics[width=\linewidth]
      {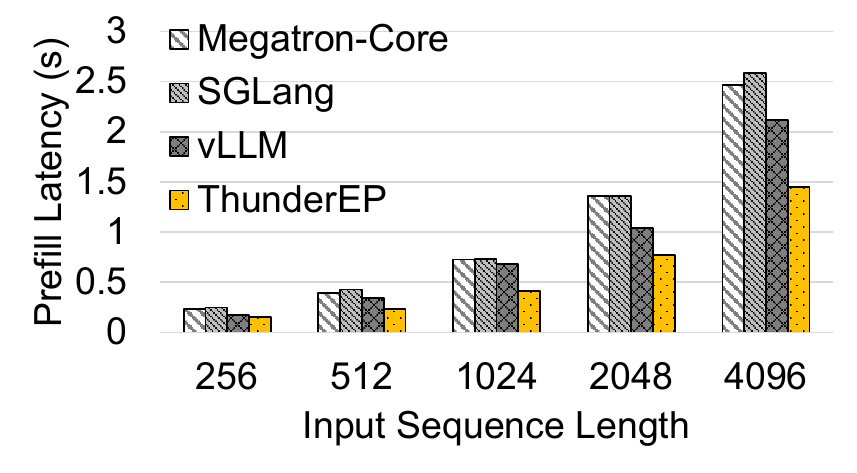}
    \caption{Qwen3-30B-A3B.}
    \label{fig:prefill-5090-qwen3}
  \end{subfigure}
  \hfill
  \begin{subfigure}[t]{0.32\textwidth}
    \centering
    \includegraphics[width=\linewidth]
      {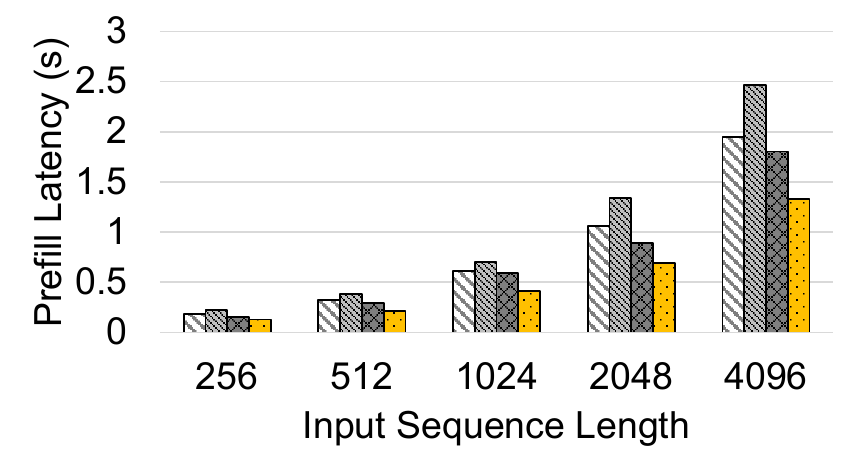}
    \caption{GPT-OSS-20B.}
    \label{fig:prefill-5090-oss-20b}
  \end{subfigure}
  \hfill
  \begin{subfigure}[t]{0.32\textwidth}
    \centering
    \includegraphics[width=\linewidth]
      {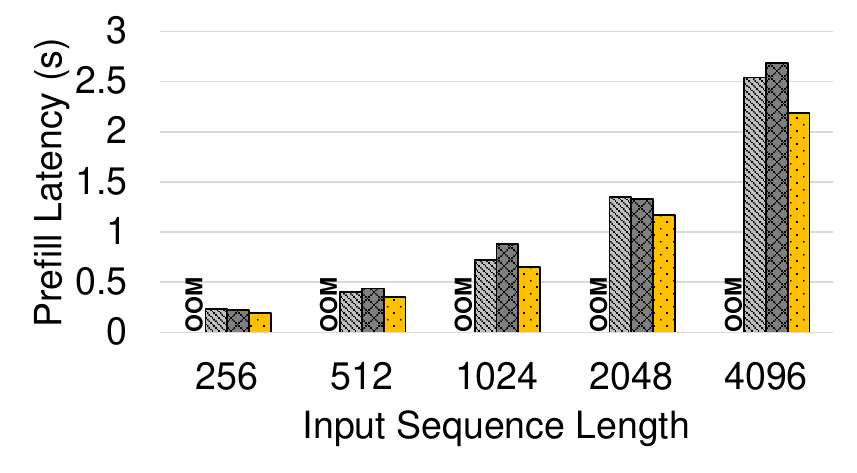}
    \caption{GPT-OSS-120B.}
    \label{fig:prefill-5090-oss-120b}
  \end{subfigure}
  \caption{Comparison of prefill performance across different input sequence
  lengths.}
  \label{fig:prefill-5090}
\end{figure*}

\begin{figure*}[t]
  \centering
  \begin{subfigure}[t]{0.32\textwidth}
    \centering
    \includegraphics[width=\linewidth]
      {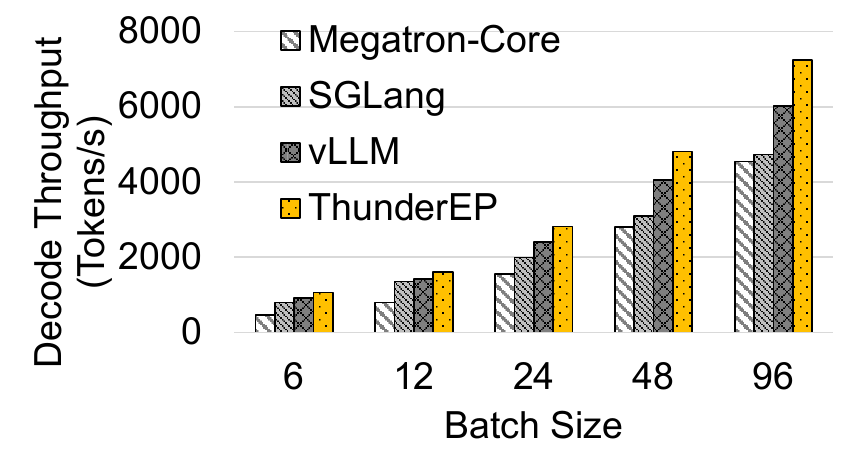}
    \caption{Qwen3-30B-A3B.}
    \label{fig:decode-5090-qwen3}
  \end{subfigure}
  \hfill
  \begin{subfigure}[t]{0.32\textwidth}
    \centering
    \includegraphics[width=\linewidth]
      {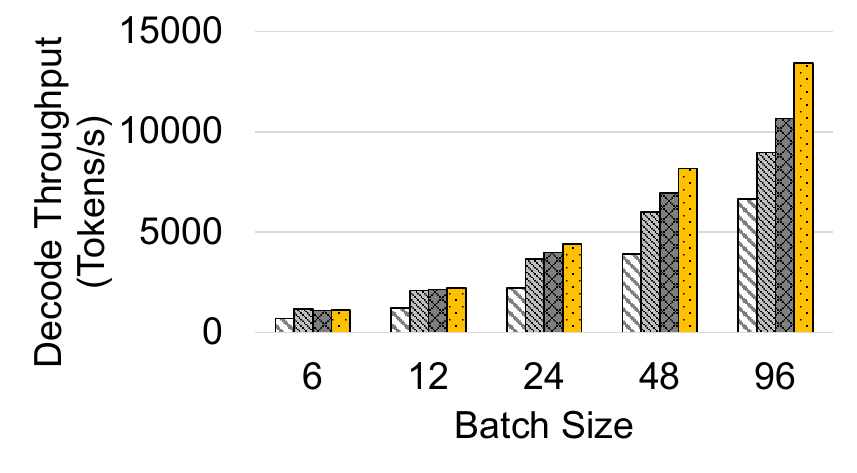}
    \caption{GPT-OSS-20B.}
    \label{fig:decode-5090-oss-20b}
  \end{subfigure}
  \hfill
  \begin{subfigure}[t]{0.32\textwidth}
    \centering
    \includegraphics[width=\linewidth]
      {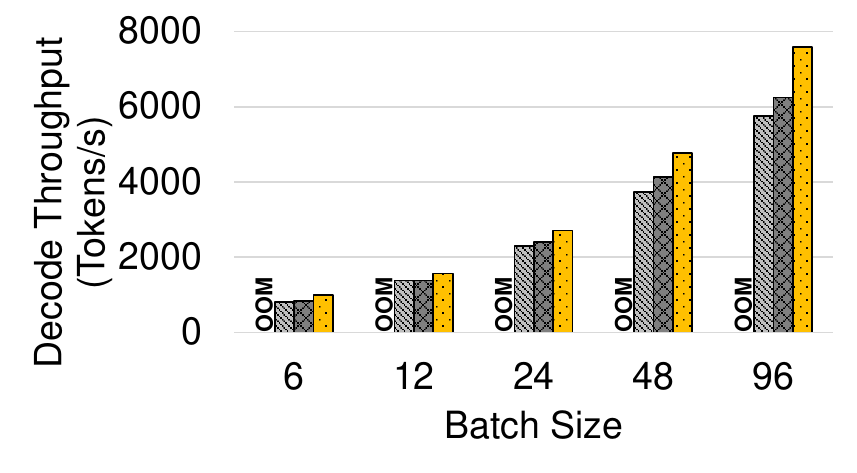}
    \caption{GPT-OSS-120B.}
    \label{fig:decode-5090-oss-120b}
  \end{subfigure}
  \caption{Comparison of decode performance across different batch sizes.}
  \label{fig:decode-5090}
\end{figure*}

\paragraph{Combine.}
Figure~\ref{fig:rs-perf} compares the Reduce-Scatter performance for token combine.
Across the 1KB--1MB range, ThunderEP achieves average latency speedups of 1.91$\times$ and 2.14$\times$ on the RTX~4090 and RTX~5090 systems, respectively.
Unlike dispatch operation, NCCL and ThunderEP move the same analytic payload volume during Reduce-Scatter. The latency improvement therefore comes primarily from the shorter dependency chain. ThunderEP transfers its contribution for each destination directly
to that destination, whereas NCCL forwards partial outputs around the ring and synchronizes between successive steps.
Avoiding these sequential relays reduces synchronization counts through host memory and shortens the critical path for small messages.

The synchronization overhead is amortized as the message size increases.
Consequently, ThunderEP and NCCL achieve similar bandwidth for large messages.
At 1GB, ThunderEP and NCCL reach 21.0 GB/s and 21.3 GB/s on the RTX~4090 system, and 24.5 GB/s and 23.2 GB/s on the RTX~5090 system.
This result also supports the dispatch performance analysis because the bandwidth advantage largely disappears when the analytic traffic volume remains unchanged. The consistent trends across both PCIe-based consumer GPU systems further show that although conventional ring algorithm perform well on systems with direct GPU access, their multi-step relays become a substantial source of overhead when every data must be transferred via a bounce buffer in CPU memory.

\subsection{End-to-end Performance}

\paragraph{Overall speedup} 

Figures~\ref{fig:prefill-5090} and~\ref{fig:decode-5090} compare end-to-end prefill latency and decode throughput on the RTX~5090 system. We include the experimental results for the RTX 4090 system in the supplementary material due to space constraints. For prefill, we use a batch size of 24 and vary the prompt length from 256 to 4K. For decode, we fix the prompt length at 4K and vary the batch size from 6 to 96. We generate 256 output tokens in all runs, which does not affect prefill latency and provides enough decoding steps for a stable throughput measurement.

ThunderEP achieves the lowest prefill latency for every model and
prompt length. Compared with vLLM, it provides average speedups of
1.42$\times$, 1.33$\times$, and 1.22$\times$ for Qwen3-30B-A3B,
GPT-OSS-20B, and GPT-OSS-120B, respectively, with a maximum speedup
of 1.66$\times$. The smaller speedup for GPT-OSS-120B arises because distributing each
shard's tokens across more local experts increases aggregate padding,
partially offsetting the communication gains. Averaged across the tested configurations, ThunderEP
achieves prefill speedups of 1.58$\times$ and 1.59$\times$ over SGLang and Megatron-Core, respectively.

For decode, ThunderEP improves throughput over vLLM by an average of
1.17$\times$, 1.12$\times$, and 1.16$\times$ for the three models,
respectively, reaching up to 1.26$\times$. Its average throughput
speedups over SGLang and Megatron-Core are 1.28$\times$ and
1.90$\times$, respectively. Megatron-Core runs out of memory (OOM) on GPT-OSS-120B because it lacks native MXFP4 computation kernels and therefore requires the model weights to be converted to BF16, increasing the memory footprint.

\begin{figure}[t]
\centering
\begin{subfigure}[b]{0.85\linewidth}
\centering
\begin{minipage}{\linewidth}
  \includegraphics[width=\linewidth]{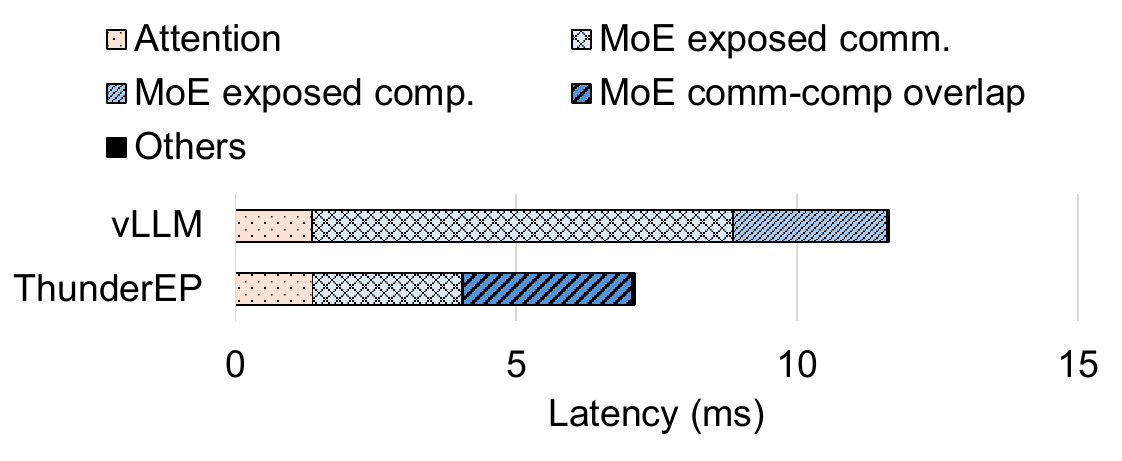}
\end{minipage}
\caption{Prefill}
\label{fig:breakdown-prefill}
\end{subfigure}
\hfill
\begin{subfigure}[b]{0.85\linewidth}
\centering
\begin{minipage}{\linewidth}
  \includegraphics[width=\linewidth]{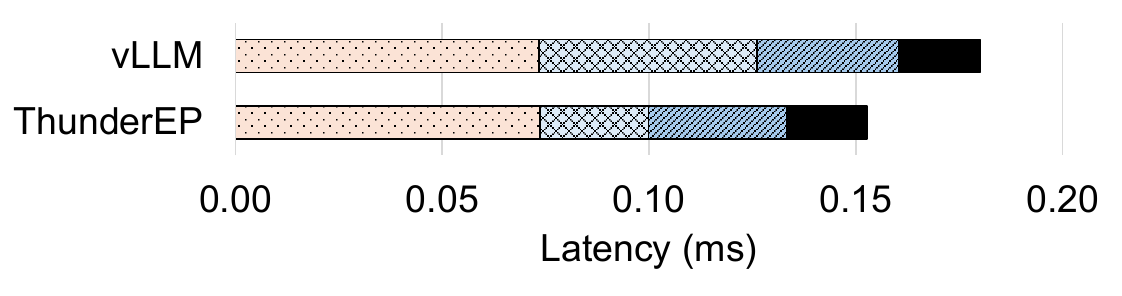}
\end{minipage}
\caption{Decode}
\label{fig:breakdown-decode}
\end{subfigure}
\caption{Latency breakdown of a Transformer layer.}
\label{fig:breakdown}
\end{figure}

\paragraph{Performance breakdown}

Figure~\ref{fig:breakdown} shows the forward latency
breakdown of a Transformer layer for Qwen3-30B-A3B on the RTX~5090
system, using a
batch size of 24 and a prompt length of 4,096. During prefill, ThunderEP's communication designs reduce
communication time compared with NCCL-based vLLM baseline. Overlapping part of the remaining transfers with expert computation provides an additional reduction in exposed latency. Together, ThunderEP reduces the exposed MoE communication latency by 64.4\% and the total Transformer layer latency by 38.9\%. Only a small fraction of expert computation remains exposed, while attention and other operations remain nearly unchanged.

During decode, compared with vLLM, ThunderEP reduces the exposed MoE communication latency by
50.1\%, lowering the Transformer layer latency by 15.2\%. Communication and expert computation are not overlapped in decode phase because each GPU processes too few tokens to form an effective pipeline. Dividing these tokens into smaller groups would add kernel launches and synchronization without providing enough computation to hide the transfers. The decode improvement therefore comes primarily from lower communication latency with the proposed single-step algorithm with efficient synchronization strategy.

\begin{table}[t]
\centering
\small
\caption{Speedup over vLLM with real-world prompts.}
\label{tab}
\vspace{-0.5em}
\begin{tabular}{l|ccc}
\hline
Datasets & Synthetic & ShareGPT & LMSYS-Chat-1M \\
\hline\hline
Prefill speedup & 1.42$\times$ & 1.38$\times$ & 1.38$\times$ \\
\hline
Decode speedup & 1.16$\times$ & 1.13$\times$ & 1.14$\times$ \\
\hline
\end{tabular}
\end{table}

\paragraph{Real-world datasets.}

Table~\ref{tab} summarizes ThunderEP's average speedup over vLLM for Qwen3-30B-A3B with prompts sampled from real-world LLM conversation datasets~\cite{sharegpt, zheng2024lmsys} on the RTX~5090 system. We truncate them to the target input length and generate the target number of output tokens while ignoring EOS tokens. We use a batch size of 24 and input lengths from 256 to 4K for prefill, and an input length of 4K and batch sizes from 6 to 96 for decode.

ThunderEP achieves a prefill speedup of 1.38$\times$ on both
datasets and decode speedups of 1.13$\times$ and 1.14$\times$ on
ShareGPT and LMSYS-Chat-1M, respectively. These gains closely
match those in the synthetic workload experiments, showing that ThunderEP remains effective with real conversation prompts.

\subsection{Ablation Study}
\label{sec:ablation}

Table~\ref{tab:ablation-main} reports an ablation study of the proposed methods applied in ThunderEP. We run Qwen3-30B-A3B on the RTX 5090 system
with an input sequence length of 4K, an output sequence length of
256, and a batch size of 24. Each row adds optimizations to the
preceding configuration. We use TTFT metric to measure prefill performance
and TPOT to measure decode performance.

\paragraph{Algorithm and synchronization.}
Replacing NCCL's ring algorithm with the proposed single-step algorithm improves
prefill performance by 1.15$\times$. Let $S$ denote the amount of input
data initially held by each of the $N$ GPUs. NCCL's ring All-Gather
moves $2(N-1)S$ bytes per GPU, whereas ThunderEP moves $NS$ bytes.
At $N=2$, both algorithms transfer the same amount of data and provide
similar performance. With 8~MB of input data per GPU, the speedup grows
to 1.26$\times$ at $N=4$ and 1.58$\times$ at $N=6$. Reduce-Scatter moves the same analytic volume in both implementations, although ThunderEP removes the sequential relays of the ring.

Combining the single-step algorithm with ThunderEP's synchronization
scheme gives a 1.15$\times$ decode speedup over vLLM. Each data
transfer in decode phase carries only a small message, so synchronization
rather than data movement dominates its latency. On six GPUs, the
single-step path reduces the number of completion checks from five to two compared with NCCL's multi-step ring.

\begin{table}[t]
  \centering
  \small
  \setlength{\tabcolsep}{4.5pt}
  \caption{Ablation study of the proposed method.}
  \label{tab:ablation-main}
  \vspace{-0.5em}
  \begin{tabular}{l|rr|rr}
    \hline
    \multirow{2}{*}{Method}
      & TTFT & TPOT & \multicolumn{2}{c}{Speedup} \\
      & (ms) & (ms) & Prefill & Decode \\
    \hline\hline

    Baseline (vLLM)
      & 2115.9 & 10.10 & $1.00\times$ & $1.00\times$ \\

    \shortstack[l]{+ Algorithm (\S~\ref{sec:algorithm}) and\\
                   \phantom{+ }synchronization
                   (\S~\ref{sec:sync})}
      & 1844.7 & 8.77 & $1.15\times$ & $1.15\times$ \\

    + DMA transfer (\S~\ref{sec:dma})
      & 1722.8 & 8.78 & $1.23\times$ & $1.15\times$ \\

    + DMA-based overlap (\S~\ref{sec:implementation})
      & 1436.4 & 8.78 & $1.47\times$ & $1.15\times$ \\
    \hline
  \end{tabular}
\end{table}

\paragraph{DMA transfer and computation overlap.}
Using the DMA engines provides a further 1.07$\times$ improvement in
prefill performance. DMA and SM transfers provide similar bandwidth
when executed alone. ThunderEP's DMA path improves performance because the transfers 
no longer share the SMs with the polling kernels that wait on peers.

Overlapping communication with expert computation adds another
1.20$\times$ improvement and raises the overall prefill speedup to
1.47$\times$. The fine-grained synchronization described in
Section~\ref{sec:sync} makes each GPU's shard available independently.
Computation on its tokens can begin as soon as the shard arrives, while
transfers from other GPUs continue.
Using the DMA engines prevents these remaining transfers from competing
with the expert GEMMs for SM resources.

On consumer GPUs, where EP communication libraries such as DeepEP cannot operate, vLLM invokes NCCL through PyTorch's
\texttt{torch.distributed} interface and schedules expert computation
only after each collective completes, providing no
communication-computation overlap. To isolate the benefit of
ThunderEP's fine-grained synchronization, we modify vLLM to provide
such overlap using the same number of pipeline units as ThunderEP. The
modified baseline invokes NCCL collectives using \texttt{async\_op} and
delays each wait until its output is needed, allowing communication for
a later unit to overlap with computation for an earlier one. Even with
this added overlap, ThunderEP achieves a 1.24$\times$ prefill speedup compared to the modified vLLM baseline.
NCCL exposes each collective's output only after data from every rank
has arrived, whereas ThunderEP can begin computation as soon as data
from each sender arrives.

\section{Conclusion}
\label{sec:conclusion}

We introduce ThunderEP, a communication design for MoE inference on
PCIe-based consumer GPU systems without P2P access. ThunderEP treats
host memory as a shared communication medium rather than simply a
staging buffer between GPUs. Its single-step All-Gather and
Reduce-Scatter algorithms remove repeated PCIe relays, DMA engine
transfers keep bulk data movement from competing with expert
computation for SM resources, and fine-grained synchronization avoids
waiting for unrelated GPUs. 
Across two PCIe systems equipped with two recent generations of consumer GPUs,
ThunderEP improves dispatch and combine communication performance over
NCCL by 2.00$\times$ and 1.53$\times$, respectively. For end-to-end
inference, ThunderEP reduces prefill latency by up to 1.66$\times$ and
increases decode throughput by up to 1.26$\times$ over vLLM, a state-of-the-art LLM serving engine.
By addressing the communication bottlenecks specific to consumer GPUs,
ThunderEP broadens the range of systems on which large MoE models can
be served efficiently.

\begin{acks}
This work was partially supported by the National Research Foundation of Korea (NRF) under Grant No. RS-2023-00222663 (Center for Optimizing Hyperscale AI Models and Platforms) and under Grant No. A400-20260031, and by the Institute for Information and Communications Technology Promotion (IITP) under Grant No. 2018-0-00581 (CUDA Programming Environment for FPGA Clusters) and No. RS-2025-02304554 (Efficient and Scalable Framework for AI Heterogeneous Cluster Systems), all funded by the Ministry of Science and ICT (MSIT) of Korea. It was also partially supported by the Korea Health Industry Development Institute (KHIDI) under Grant No. RS-2025-25454559 (Frailty Risk Assessment and Intervention Leveraging Multimodal Intelligence for Networked Deployment in Community Care), funded by the Ministry of Health and Welfare (MOHW) of Korea. Additional support was provided by the BK21 Plus Program for Innovative Data Science Talent Education (Department of Data Science, Seoul National University, No. 5199990914569) and the BK21 FOUR Program for Intelligent Computing (Department of Computer Science and Engineering, Seoul National University, No. 4199990214639), both funded by the Ministry of Education (MOE) of Korea. This work was also partially supported by the Advanced GPU Utilization Support Program, funded by the Ministry of Science and ICT (MSIT) of Korea and operated by the National IT Industry Promotion Agency (NIPA). Research facilities were provided by the Institute of Computer Technology (ICT) at Seoul National University.
\end{acks}

\bibliographystyle{ACM-Reference-Format}
\bibliography{reference}

\clearpage
\onecolumn
\appendix

\section{Additional Results}
\label{sec:appendix}

\begin{figure}[ht]
  \centering

  \begin{subfigure}[t]{0.32\textwidth}
    \centering
    \includegraphics[width=\linewidth]
      {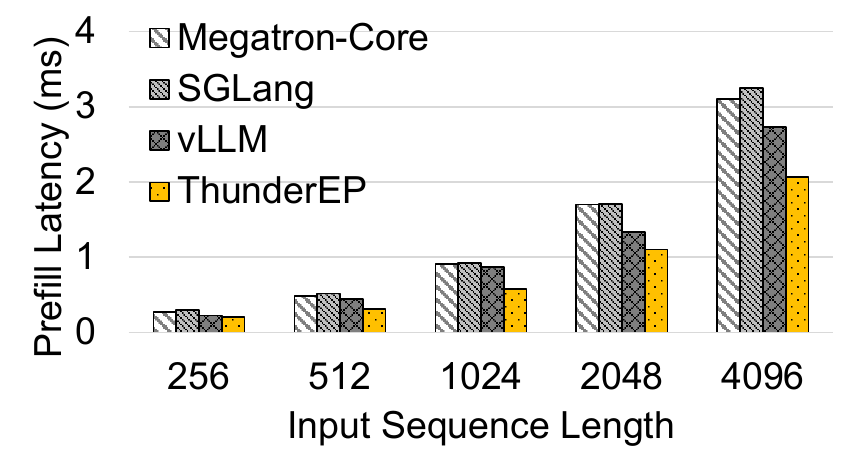}
    \caption{Qwen3-30B-A3B.}
    \label{fig:prefill-4090-qwen3}
  \end{subfigure}%
  \hspace{0.03\textwidth}%
  \begin{subfigure}[t]{0.32\textwidth}
    \centering
    \includegraphics[width=\linewidth]
      {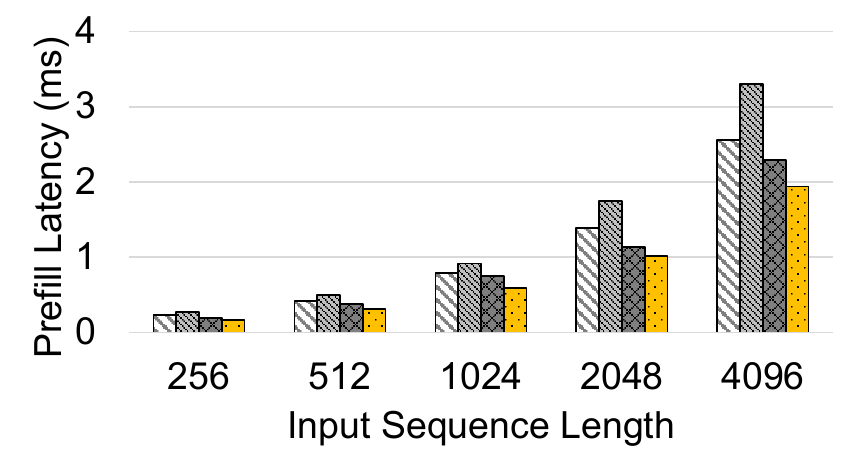}
    \caption{GPT-OSS-20B.}
    \label{fig:prefill-4090-oss-20b}
  \end{subfigure}

  \caption{Comparison of prefill performance across different input
  sequence lengths on the RTX 4090 system. We use a batch size of 24 and 256 output tokens, and vary the prompt length from 256 to 4K. We exclude GPT-OSS-120B since all baselines run GPU OOM.}
  \label{fig:prefill-4090}

  \vspace{1em}

  \begin{subfigure}[t]{0.32\textwidth}
    \centering
    \includegraphics[width=\linewidth]
      {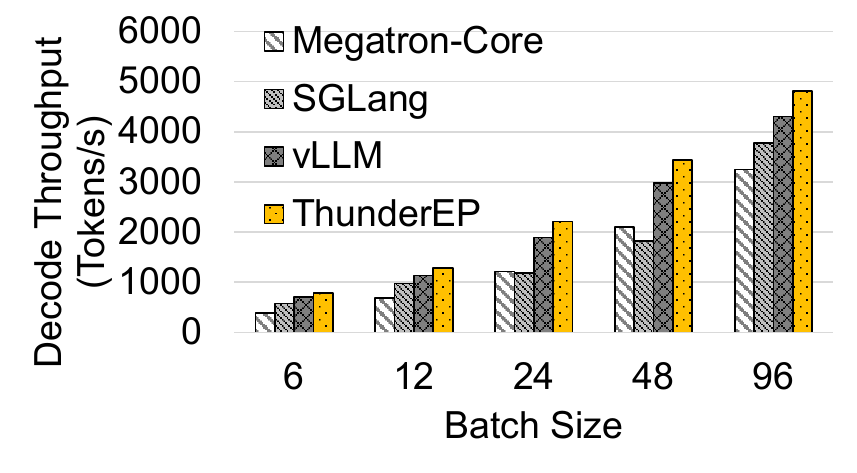}
    \caption{Qwen3-30B-A3B.}
    \label{fig:decode-4090-qwen3}
  \end{subfigure}%
  \hspace{0.03\textwidth}%
  \begin{subfigure}[t]{0.32\textwidth}
    \centering
    \includegraphics[width=\linewidth]
      {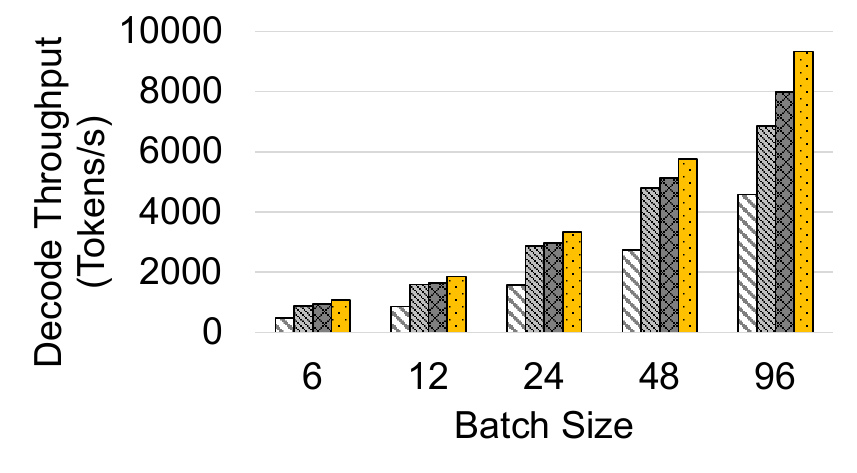}
    \caption{GPT-OSS-20B.}
    \label{fig:decode-4090-oss-20b}
  \end{subfigure}

  \caption{Comparison of decode performance across different batch sizes
  on the RTX 4090 system. We use a prompt length of 4K and 256 output tokens, and vary the batch size from 6 to 96. We exclude GPT-OSS-120B since all baselines run GPU OOM.}
  \label{fig:decode-4090}
\end{figure}

\end{document}